\documentclass[prd,aps,twocolumn,superscriptaddress,floatfix,preprintnumbers,nofootinbib]{revtex4-2}
\usepackage{amsmath}
\usepackage{graphicx}
\usepackage{xcolor}
\usepackage{amsfonts}
\usepackage{natbib}
\usepackage{tikz}
\usepackage{siunitx}
\usepackage{tikz}
\usetikzlibrary{math}
\usepackage{amssymb,amsmath,bm}
\usepackage[citecolor=blue,pdfa=true,linktocpage=true,urlcolor=blue,colorlinks=true]{hyperref}
\usepackage[export]{adjustbox}
\usepackage{multirow}
\usepackage{physics}
\usepackage[capitalise]{cleveref}
\usepackage{soul}
\creflabelformat{equation}{#2\textup{#1}#3}

\newcommand{\aap}{{Astron.~Astrophys.}}
\newcommand{\apjl}{{Astrophys.~J.~Lett.}}
\newcommand{\apjs}{{Astrophys.~J.~Supp.}}

\newcommand{\mnras}{{Mon.~Not.~R.~Astron.~Soc.}}
\newcommand{\jcap}{JCAP}

\newcommand{\be}{\begin{equation}}
\newcommand{\ee}{\end{equation}}

\newcommand{\bea}{\begin{eqnarray}}
\newcommand{\eea}{\end{eqnarray}}
\newcommand{\bdm}{\begin{displaymath}}
\newcommand{\edm}{\end{displaymath}}

\def\fNL{f_{\mathrm{NL}}}

\newcommand{\bs}{\mathbf{s}}
\newcommand{\bk}{\mathbf{k}}
\newcommand{\bn}{\mathbf{n}}

\def\bx{\bold{x}}
\def\bv{\bold{v}}
\def\bq{\bold{q}}

\def\bn{\bold{n}}
\def\mH{\mathcal{H}}

\newcommand{\HL}[1]{{\color{black}#1}}

\begin{document}

\title{Linear Relativistic Corrections in the  Spherical Fourier-Bessel Power Spectrum}

\author{Robin Y. Wen}
\email{ywen@caltech.edu}

\affiliation{California Institute of Technology, 1200 East California Boulevard, Pasadena, California 91125, USA}

\author{Henry S. Grasshorn Gebhardt}
\affiliation{California Institute of Technology, 1200 East California Boulevard, Pasadena, California 91125, USA}
\affiliation{Jet Propulsion Laboratory, California Institute of Technology, Pasadena, California 91109, USA}

\author{Chen Heinrich}
\affiliation{California Institute of Technology, 1200 East California Boulevard, Pasadena, California 91125, USA}

\author{Olivier Dor\'e}
\affiliation{California Institute of Technology, 1200 East California Boulevard, Pasadena, California 91125, USA}
\affiliation{Jet Propulsion Laboratory, California Institute of Technology, Pasadena, California 91109, USA}

\begin{abstract}
The three-dimensional galaxy power spectrum is a powerful probe of primordial non-Gaussianity and additional general relativistic (GR) effects on large scales, which can be constrained by the current and upcoming large-scale structure surveys. In this work, we calculate the linear-order relativistic power spectrum in the spherical Fourier-Bessel (SFB) basis, a coordinate system that preserves the geometry of the curved sky and fully accounts for the wide-angle effect. In particular, we model the GR effects present in the discrete SFB power spectrum, which is a more efficient and stable decomposition of the galaxy density field compared to the continuous SFB basis in the presence of radial windows. To validate our GR calculations, we introduce a mapping between the angular power spectrum and the SFB power spectrum, and we compare our calculations with outputs from \texttt{CLASS}. We discuss the rich pattern of GR effects in the SFB basis and compare the GR effects to the local primordial non-Gaussianity (PNG) effect. The Doppler and lensing effects have different angular and Fourier dependence compared to the PNG in the SFB basis, while the gravitational potential term is more degenerate with the PNG and comparable to a signal of $f_{\rm NL}\sim 1$. We also discuss the potential opportunities of extracting the lensing effect through SFB modes in upcoming LSS surveys.
\end{abstract}

\maketitle
\tableofcontents

\newpage
\section{Introduction}
Current and upcoming surveys such as DESI \cite{16DESI}, Euclid \cite{16Euclid}, LSST \cite{19LSST}, SPHEREx \cite{14SPHEREx}, and Roman \cite{20Eifler_Roman} will trace the large-scale structure (LSS) of the Universe over increasingly large volumes by measuring the positions and spectra of hundreds of millions of galaxies. The unprecedented constraining power of these surveys offers exciting prospects for probing fundamental physics such as the nature of inflation, dark matter, and dark energy.

To fully leverage the potential of these surveys, it is crucial to model the observed data accurately. In practice, all tracers of the LSS are observed via photon arrival directions (right ascension and declination) and redshifts, inferred from the shift in frequency of the observed spectral energy distributions of the galaxy relative to the rest-frame frequency. That is, we do not have access to the rest-frame galaxy density (or line intensity) directly. Hence, an essential ingredient in the interpretation
of large-scale structure is the modeling of redshift space distortion (RSD) effects -- the mapping between the observed coordinates of an object on the sky and its true position on the rest-frame light cone.

Traditionally, the Newtonian modeling of the RSD only considers the impacts of the galaxies' peculiar velocities on the observed redshift. As we probe galaxy clustering at increasingly larger scales, such Newtonian modeling of the RSD breaks down when general relativity (GR) becomes important. The GR corrections present in the galaxy density fluctuation scale with integer powers of $\mH/k$, where $\mH$ is the conformal Hubble factor, and $k$ is the comoving 3D Fourier mode, so they can be safely ignored in the previous generation surveys with smaller volume. However, the relativistic corrections will become non-negligible when the current and upcoming LSS surveys start to probe cosmological modes whose length scale is comparable to the Hubble scale. 

These GR effects contain additional cosmological information on the cosmic velocity field through the Doppler effect, the gravitational potential field through the (integrated) Sachs-Wolfe effect and Shapiro time-delay effects, and the perturbations of the geometry through gravitational lensing. Since Einstein’s equations specify the relationships among the density, velocity, and potential perturbations, these relativistic effects allow us to test general relativity and constrain cosmological parameters. Modeling relativistic effects is also critical for galaxy surveys aiming to constrain the local primordial non-Gaussianity (PNG) to the precision of $\sigma(f_{\rm NL})\sim1$ \cite{23WAGR,23Foglieni,15Alonso_GR_single,15Alonso_GR_multi}, since local PNG is constrained at the largest scales where the GR corrections become non-negligible theoretical contaminants.

The linear-order GR effects have already been calculated for the angular power spectrum~\cite{11ChallinorLPS,11Bonvin_GR,13DiDio_classgal,18Schoneberg_cl_fftlog_gr} (also known as tomographic spherical harmonics, abbreviated as TSH), and incorporated into the well-established linear Boltzmann codes \texttt{CAMB} \cite{CAMB} and \texttt{CLASS} \cite{11class} with accurate implementation. However, for other common two-point statistics, in particular for those often used in spectroscopic surveys such as two-point correlation functions (2CF) and power spectrum multipoles (PSM), a fully consistent calculation of linear-order GR effects has only been carried out in the past few years. Refs.~\cite{18Tansella-GR} and \cite{22CatorinaGR-P} have achieved the numerical evaluation of linear-order GR effects in both configuration-space 2CF and Fourier-space PSM, culminating in the public codes \texttt{COFFE} \cite{18Tansella_coffe} and \texttt{GaPSE} \cite{23Foglieni}.

Both correlation function multipoles (2CFM) and power spectrum multipoles are statistics in Cartesian coordinates, and one needs to explicitly model the wide-angle (WA) effects \cite{15YooWA,16ReimbergWA,18CastorinaWA,23WAGR,24Benabou_WA}, the fact that the commonly assumed plane-parallel approximation breaks down for galaxy pairs with large separations in a wide-field survey, in order to perform a consistent calculation of GR effects. The Cartesian Fourier basis consists of the eigenfunctions of the Laplacian in Cartesian coordinates, while in contrast, the spherical Fourier-Bessel (SFB) basis, consisting of the eigenfunctions of the Laplacian in spherical coordinates, is a more natural basis for analyzing the curved sky and addressing the WA effects.

Besides preserving the geometry of the
curved sky, the SFB power spectrum also accounts for the differing lines-of-sight (LOS) of the galaxies in a galaxy pair and contains the full information on the anisotropic clustering and redshift evolution, while in contrast the Cartesian-based 2CFM and PSM only adopt a single LOS for a galaxy pair and erase the redshift evolution information within the redshift bin. Furthermore, Ref.~\cite{24PSM_SFB} has shown the PSM as a compression of the SFB power spectrum and demonstrated the use of SFB basis for the ultra-large modeling of both the signals and the covariance in the standard Cartesian analysis. In addition, the SFB basis allows one to cross-correlate a 3D density field (with both angular and redshift information) with a projected density field (with only angular information) while retaining the radial modes provided by spectroscopic redshift. It is therefore the natural and optimal basis
for cross-correlating spectroscopic surveys with photometric galaxy, cosmic shear, and CMB lensing surveys \cite{21Zhang_SFB_TSH,17Passaglia_2D_3D}. It is for all these reasons that the SFB analysis has gained significant traction in recent years \cite{12Leistedt,13Yoo_GR_SFB,13Pratten_SFB,14Nicola_SFB,15Lanusse_SFB,16SFB_IM,17Passaglia_2D_3D,19Samushia_SFB,20Wang_Hybrid_Estimator,21Zhang_SFB_TSH,21Gebhardt_SuperFab,22Khek_SFB_fast,23Gebhard_SFB_eBOSS,23Benabou_SFB,24PSM_SFB,24Semenzato_SFB_GR}. 

Despite these advantages, the theoretical computation of the SFB power spectrum is less developed compared to the PSM, especially at the mildly non-linear scales $(k\geq 0.05\,h/{\rm Mpc)}$ where the 1-loop correction becomes important. Even at the linear level, the numerical evaluation of the SFB power spectrum (PS) remains non-trivial due to its three-dimensional nature involving a large number of modes. Recently, Ref.~\cite{23Gebhard_SFB_eBOSS} has employed an efficient method of calculating the SFB PS at the linear level under the Newtonian RSD. Under their method, the numerical evaluation of the SFB only takes seconds on a single CPU, enabling Ref.~\cite{23Gebhard_SFB_eBOSS} to perform a Markov chain Monte Carlo (MCMC) analysis on mocks with the SFB PS. 

Preparing for the upcoming cosmological data that have the potential of measuring GR effects and achieving stringent constraints on $f_{\rm NL}$, it is critical to compute the GR effects with efficiency and accuracy in the two-point statistics. In this work, we build upon the method introduced in Ref.~\cite{23Gebhard_SFB_eBOSS} to calculate the linear-order general relativistic corrections present in the SFB power spectrum. Most of the literature about the theoretical modeling of the SFB PS, such as Refs.~\cite{14Nicola_SFB,15Lanusse_SFB,16SFB_IM,20Wang_Hybrid_Estimator,22Khek_SFB_fast}, only contains the linear Newtonian RSD effects. Ref.~\cite{24PSM_SFB} further adds the Doppler effect, while Ref.~\cite{21Zhang_SFB_TSH} considers the lensing effect. To our knowledge, Refs.~\cite{13Yoo_GR_SFB} and \cite{24Semenzato_SFB_GR} are the only works that have calculated all the linear-order GR effects in the SFB PS, but they adopt the continuous SFB basis. Compared to these two works, we will calculate the GR effects in the discrete SFB basis instead of the continuous one, since the discrete SFB basis is a more efficient and stable decomposition of the galaxy density field and a more natural choice for numerical computation and data estimation. In addition, we will place greater emphasis on the numerical technique compared to the previous works and provide explicit validation for our calculations.

This paper is outlined as follows: we first summarize the GR effects at linear order in \cref{sec:GR}. In particular, we will clarify the meaning of Newtonian approximations in \cref{sec:Newtonian_approx} and provide a review of the local PNG in the relativistic context in \cref{sec:PNG}. We next outline the SFB formalism in \cref{sec:SFB}: we will review the basics of the SFB decomposition (\cref{sec:SFB-decomp}) and emphasize the benefit of discrete SFB basis over the continuous one (\cref{sec:discrete-benefit}). We present the analytical formalism for computing the SFB power spectrum and GR effects in \cref{sec:compute-analytic}. Our main numerical approach is based on the Iso-qr integration method proposed in Ref.~\cite{23Gebhard_SFB_eBOSS}, which we will review and extend for GR effects in \cref{sec:compute-numerics}. We present and discuss our numerical results in \cref{sec:result} and outline the future prospects in \cref{sec:conclusion}.

We defer some technical details and extensions of our work to the Appendices. We outline the mappings between the SFB power spectrum and other two-point statistics in \cref{sec:SFB-map}. In particular, the mappings between the SFB PS and the angular power spectrum developed in \cref{sec:TSH-SFB} are critical for developing the theoretical formalism of SFB and validating \HL{our implementation of the GR effects}. We next review the derivation of the angular kernels for all GR effects in \cref{sec:TSH-decomp}. In \cref{sec:validation}, we provide validation for our GR effects computation by comparing our results with outputs from \texttt{CLASS}. We then discuss approximation schemes for the lensing effect (\cref{sec:lens-approx}) and the divergence encountered in the SFB monopole (\cref{sec:divergence}). We last provide details for the Iso-qr integration scheme introduced in Ref.~\cite{23Gebhard_SFB_eBOSS} and further extend the method in \cref{sec:qr_integration}.

Though focused on galaxy surveys, our work will have implications for future intensity mapping surveys where relativistic effects can also become important \cite{13Hall_GR,20SKA}. Our formalism for calculating the lensing convergence in the SFB power spectrum can also be used for lensing potential and lensing shear, which have different angular dependence compared to the lensing convergence \cite{17Lemos_limber_lensing}.

Throughout this work, we consider the simplest case of a full-sky window with uniform radial selection in some redshift bin from $z_{\rm min}$ and $z_{\rm max}$ (a top-hat radial window), that is we consider a galaxy survey with uniform galaxy density in the comoving space with a finite redshift footprint without any angular mask. The window convolution can be further modeled based on Ref.~\cite{21Gebhardt_SuperFab} where the window mixing matrix is applied on top of the full-sky SFB power spectrum. We ignore the integral constraint (also known as the local average effect) \cite{19Mattia_IC,23Gebhard_SFB_eBOSS} and the observer's terms \cite{20GrimmGR,22CatorinaGR-P} in our numerical evaluations. All plots presented in this work assume a best-fit Planck 2018 $\Lambda$CDM cosmology \cite{18Planck_Parameter}. We use \texttt{julia} for all numerical implementations, and the code will be made publicly accessible in a follow-up paper.

\section{Relativistic Effects in Observed Galaxy Number Count}\label{sec:GR}
In this section, we review the relativistic effects present in the galaxy clustering at the linear order. It is also referred to as the projection effects in some literature \cite{18Desjacques_Galaxy_Bias,23Foglieni}. The starting point is the linear-order relativistic expression for the so-called observed galaxy (source) number count. For a sample of galaxies observed with a given set of selection criteria, we denote the galaxy number density as $N_{\rm g}(\hat{\bn}, z)$ in terms of the observed angle $\hat{\bn}$ and redshift $z$, and the observed galaxy number count is
\begin{align}
\delta_{\rm g}(\hat{\bn},z)=\frac{N_{\rm g}(\hat{\bn},z)-\langle N_{\rm g}(\hat{\bn},z)\rangle}{\langle N_{\rm g}(\hat{\bn},z)\rangle},
\end{align}
where $\langle...\rangle$ is the ensemble average at fixed observed redshift. The galaxy number count $\delta_{\rm g}(\hat{\bn},z)$ is an observable that is gauge invariant, so it can be expressed in any gauge for the perturbed Friedmann-Lemaître-Robertson-Walker metric. It is mostly commonly expressed in the conformal Newtonian gauge (CNG):
\begin{align}
ds^2=a(\tau)^2[-(1+2\Psi)d\tau^2+(1-2\Phi)\delta_{ij}dx^idx^j],
\label{eq:CNG}
\end{align}
where $\tau$ denotes the conformal time, $\bx$ is the Cartesian comoving coordinate in the configuration space, and $\Psi$ and $\Phi$ are the Bardeen potentials, corresponding to the temporal and spatial metric perturbations in the CNG. \cref{eq:CNG} is named the Newtonian gauge since $\Psi$ coincides with the gravitational potential under Newtonian gravity. Here we only consider the metric perturbation caused by scalar modes in an otherwise flat background (that is $\Omega_{\rm K}=0$).

The linear relativistic calculation for the galaxy number count has been carried out in Refs.~\cite{09Yoo_GR,11ChallinorLPS,11Bonvin_GR,12JeongLPS} with the observer terms added in Refs.~\cite{20GrimmGR,18Scaccabarozzi_GR_TP,22CatorinaGR-P}. Following the notations of Refs.~\cite{23WAGR} and \cite{22CatorinaGR-P}, we have
\begin{align}
\delta_{\rm g}^{\rm rel}(\hat{\bn},z)&=b_1D_{\rm m}-\frac{1} {\mH}\frac{\partial\vec{v}}{\partial x}\cdot\hat{\bn}\nonumber\\
&-(2-5s)\kappa\nonumber\\
&-\mathcal{A}_1(\vec{v}-\vec{v}_o)\cdot\hat{\bn}+(2-5s)\vec{v}_{o}\cdot\hat{\bn}\nonumber\\
&+\mathcal{A}_1(\Psi-\Psi_o)+\left(\mathcal{A}_1\mathcal{H}_0-\frac{2-5s}{x}\right)V_{o}\nonumber\\
&\qquad-(2-5s)\Phi+\Psi+\frac{1}{\mH}\dot{\Phi}+(b_{\rm e}-3)\mathcal{H}V\nonumber\\
&-\frac{2-5s}{x}\int_{\tau_0}^{\tau(z)}(\Psi(\tau')+\Phi(\tau'))\,d\tau'\nonumber\\
&-\mathcal{A}_1\int_{\tau_0}^{\tau(z)}(\dot{\Psi}(\tau')+\dot{\Phi}(\tau'))\,d\tau'.
\label{eq:GR}
\end{align}
In the above equation, we have grouped the relativistic effects\footnote{Different works often adopt different groupings for the linear-order GR effects. For example, Ref.~\cite{22CatorinaGR-P} groups Shapiro and ISW terms together as the integrated gravitational potentials, while Ref.~\cite{23WAGR} groups Doppler and NIP terms as the non-integrated relativistic effects. \cref{eq:GR} can also be expressed in terms of perturbations on physical quantities such as redshift, luminosity distance, and volume \cite{10Yoo_GR,14Dio_GNC_2nd}.} as follows: standard density plus RSD (first line, abbreviated as DRSD), lensing (second line), Doppler (third line), non-integrated potential (fourth and fifth lines, abbreviated as NIP hereafter), Shapiro (sixth line), and integrated Sachs–Wolfe (seventh line, referred to as ISW hereafter). We refer to the NIP, Shapiro, and ISW terms (last three lines) together as the gravitational potential term (abbreviated as GP). 

We next explain the cosmological quantities used in \cref{eq:GR}. There, $D_{\rm m}$ is the gauge-invariant density contrast that coincides with the matter density fluctuation in the comoving gauge\footnote{The comoving gauge is defined such that the fluid 3-velocity is set to vanish. We use the comoving gauge for the matter density contrast here because it is the proper gauge for galaxy bias expansion (see \cref{sec:PNG}), and the matter power spectrum in the comoving gauge can be directly calculated from linear Boltzmann codes.}, and $\bv$ is the peculiar velocity of the galaxies in the CNG, assumed to follow the matter velocity. The velocity potential $V$ is related to the peculiar velocity $\bv$ via $\bv=-\nabla V$, and we define the velocity scalar $v$ through $v(k)\equiv kV(k)$ such that $v$ and $\bv$ will have the same dependence on $k$ in the Fourier space. The lensing convergence $\kappa$ is given by
\begin{align}
&\kappa(\hat{n},z)=\frac{1}{2}\nabla^2_{\hat{\bn}'}\psi^{\rm lens}\nonumber
\\
&=-\frac{1}{2}\nabla^2_{\hat{\bn}}\int_{\tau_0}^{\tau(z)}\frac{\tau'-\tau(z)}{(\tau_0-\tau(z))(\tau_0-\tau')}(\Phi(\tau')+\Psi(\tau'))\,d\tau',
\label{eq:kappa}
\end{align}
where $\psi^{\rm lens}$ is the lensing potential, and the integration occurs along the LOS $(\hat{\bn}=\hat{\bn}')$. In \cref{eq:GR}, we use the subscript $o$ to indicate that the quantity is evaluated at the observer’s location (e.g., $\Psi_o$ indicates the temporal gravitational potential at the observer), the subscript $0$ to indicate the current time (e.g. $\tau_{0}$), and the dot to refer to a partial derivative with respect to the conformal time.

We have also dropped the explicit redshift dependence for variables in \cref{eq:GR}, that is $b_1$, $\mH$, $\mathcal{A}_1$, $s$, and $b_{\rm e}$ all depend upon the source's redshift. We next give the explicit definitions of these redshift-dependent variables. The linear galaxy bias $b_1(z)$ characterizes the relationship between the dark matter and the proper galaxy density fluctuation in the comoving gauge. The conformal Hubble parameter is defined as $\mH(z)\equiv\frac{da}{d\tau}/a=aH(z)$, and it is in the natural unit assuming $c=1$. $\mathcal{A}_1$ is defined through
\begin{align}
\mathcal{A}_1(z)\equiv\frac{\dot{\mH}}{\mH^2}+\frac{2-5s}{\mH x}+5s-b_{\rm e}.
\end{align}

The evolution bias $b_{\rm e}$ measures the evolution of the galaxy comoving density:
\begin{align}
b_{\rm e}&\equiv \frac{\partial\ln \bar{n}_g}{\mH \partial \tau} = \frac{\partial\ln \bar{n}_g}{\partial\ln a},
\label{eq:be}
\end{align}
where $\bar{n}_g(x)$ denotes the mean galaxy density in the comoving space, and it can be converted to the physical mean galaxy density $\bar{n}_g^{\rm phys}$ via $\bar{n}_g=a^3\bar{n}_g^{\rm phys}$. The magnification bias $s$ measures the change in the galaxy number density with respect to the luminosity cut $L_{\rm c}$ (assuming the galaxy survey is magnitude limited) at fixed redshift:
\begin{align}
s&\equiv-\frac{2}{5}\frac{\partial \ln \bar{n}_g}{\partial \ln L}\bigg|_{\ln L_{\rm c}}.
\label{eq:s}
\end{align}

The above definition of magnification bias is based on a simple magnitude cut for the galaxy samples. For realistic surveys with more complicated sample selections, the bias $s$ measures the response of the sample selection function to the change in the observed galaxies' flux due to magnification. In the $(2-5s)$ factor present in \cref{eq:GR}, the factor $2$ characterizes how the observed galaxy number density changes with respect to the observed angular area, which varies across different LOS due to cosmological perturbations. This is often known as the geometric magnification. Besides changing the observed angular area, cosmological perturbations will also change the observed galaxies' flux from its background expectation, that is the luminosity distance at each LOS will be perturbed from the background luminosity distance. Since the survey selection of galaxy samples primarily depends on the observed flux of galaxies at different photometric bands, the flux magnification will then change whether a particular galaxy is selected or not, thereby changing the observed number count. The factor $5s$ \HL{characterizes} how the observed galaxy number density changes with respect to this flux magnification.

The evolution and magnification biases determine the amplitude of relativistic corrections in the number count, so measuring these two biases is important for upcoming LSS surveys that have the constraining power to measure GR effects or in surveys that can reach $\sigma(f_{\rm NL})\sim 1$ for which GR effects are important~ \cite{21Maartens_be,23WAGR,23Bonvin_Dipole,23Foglieni}. Note that the magnification bias is already important for current photometric surveys that probe galaxy clustering and cosmic shear \cite{22Elvin-Poole_DES_MB,24Krolewski_CMB_quadar_fnl,22Euclid_mag_photoz}. For examples of measuring magnification bias in real surveys, see Refs.~\cite{22Elvin-Poole_DES_MB} and \cite{24Wenzl_mag_bias_SDSS}.

We emphasize that the linear-order relativistic galaxy number count given in \cref{eq:GR} applies to any metric theory of gravity. For this work, we will restrict ourselves to GR under $\Lambda$CDM cosmology for all the numerical calculations. Since the perturbed Einstein equations are not used to derive \cref{eq:GR}, it still applies when one considers different dark energy models with perturbation or modified gravity models.

\subsection{Newtonian Approximations}\label{sec:Newtonian_approx}

The expression for the observed galaxy number count in \cref{eq:GR} includes all relativistic terms at linear order. However, most of these effects are not needed in the previous LSS surveys with relatively small volumes and small angular coverage focusing on scales $k\gtrapprox 0.01\,h/{\rm Mpc}$. For a spectroscopic survey with precise redshift, the linear modeling has only used the first line of \cref{eq:GR} \cite{17BOSS_fourier_wedge,21eBOSS_quasar_fnl}:
\begin{align}
\delta_{\rm g}^{\rm spec}(\hat{\bn},z)&=b_1D_{\rm m}-\frac{1} {\mH}\frac{\partial\vec{v}}{\partial x}\cdot\hat{\bn}\,.
\label{eq:spec}
\end{align}
This is the galaxy density and the standard Newtonian RSD term, which are used to calculate the two-point correlation function multipoles $\xi_{L}(s)$ and power spectrum multipoles $P_{L}(k)$. These two statistics are the standard galaxy clustering statistics used in spectroscopic surveys for extracting cosmological information, and \cref{eq:spec} has been sufficient for modeling galaxy clustering at the linear level for the Stage-III surveys.

However, as demonstrated in Refs.~\cite{24Benabou_WA,10Raccanelli_WA_sim,24PSM_SFB}, an additional velocity term needs to be included to \cref{eq:spec} for analyzing the large-scale clustering of mocks or surveys with large angular coverage. The full expression for the linear RSD in the Newtonian limit is \cite{87Kaiser,98Hamilton_RSD}:
\begin{align}
\delta_{\rm g}^{\rm Newt}(\hat{\bn},z)&=b_1D_{\rm m}-\frac{1} {\mH}\frac{\partial\vec{v}}{\partial x}\cdot\hat{\bn}-\frac{\alpha}{\mathcal{H}x}\vec{v}\cdot\hat{\bn}\,,
\label{eq:Newtonian}
\end{align}
in which the third term is generated from the Jacobian associated with the change of coordinates caused by the Newtonian RSD. This additional velocity term is often referred to as the $\alpha$-term \cite{18CastorinaWA} or the Newtonian Doppler term \cite{24Benabou_WA}. In the plane-parallel limit, this additional velocity term becomes negligible as $x\to\infty$, which reduces \cref{eq:Newtonian} to \cref{eq:spec} used in Stage-III spectroscopic surveys. When one analyzes the large-scale clustering under the Newtonian RSD in the full-sky limit where wide-angle effects are also important, the full Newtonian expression in \cref{eq:Newtonian} should be adopted.

One can see that the $\alpha$-term has the same form as the Doppler term in the GR number count (the third line of \cref{eq:GR}) albeit with a different coefficient. The coefficient $\alpha$ is defined as \cite{15YooWA,18CastorinaWA,24Benabou_WA}:
\begin{align}
\alpha(x) \equiv \frac{\partial \ln[x^2\bar{n}_g(x)]}{\partial \ln x}=2-x\mH b_{\rm e}
\label{eq:alpha},
\end{align}
and it is related to the evolution bias defined in \cref{eq:be}. In the absence of the evolution bias, that is the galaxy has a constant density in the comoving space, $\alpha=2$.

Comparing the coefficients for the Newtonian and GR Doppler terms, we see 
\begin{align}
\mathcal{A}_1-\frac{\alpha}{\mathcal{H}x}=\frac{\dot{\mH}}{\mH^2}-\frac{5s}{\mH x}+5s.
\label{eq:GR-Newt_Doppler}
\end{align}
Under GR, the peculiar velocity not only impacts the observed redshift but also changes the luminosity distance and thereby the observed flux of the galaxy, which adds terms proportional to the magnification bias~\cite{15YooWA}. Furthermore, the redshift space distortion takes place through the past light cone under GR and causes not only spatial but also time displacement, which explains the $\dot{\mH}/\mH^2$ term present in \cref{eq:GR-Newt_Doppler}. Both of these effects are ignored in the Newtonian modeling of RSD, which explains the difference of the coefficients between the Newtonian and GR Doppler effects\footnote{See Sec.~2.4 of Ref.~\cite{15YooWA} for a more detailed discussion on comparing the Newtonian Doppler with the GR Doppler term.}.

For modeling the galaxy clustering observed in photometric surveys, it is standard to use \cite{20Fang_FFTlog,22DESY3,24Krolewski_CMB_quadar_fnl}
\begin{align}
\delta_{\rm g}^{\rm photo}(\hat{\bn},z)&=b_1D_{\rm m}-\frac{1} {\mH}\frac{\partial\vec{v}}{\partial x}\cdot\hat{\bn}-(2-5s)\kappa,
\label{eq:photo}
\end{align}
which adds the lensing term compared to \cref{eq:spec} used for spectroscopic surveys. Due to the presence of substantial redshift errors in photometric surveys, it is standard to use the angular correlation function $\omega(\theta)$ or the angular power spectrum $C_{\ell}$ to probe the radially-projected galaxy clustering. For the 3D clustering often probed by power spectrum multipoles $P_{L}(k)$, the lensing term only becomes important for ultra-large scales at lower $k$~\cite{22CatorinaGR-P,23WAGR}, which justifies the ignorance of the lensing term in the Stage-III spectroscopic surveys focusing on $k\gtrapprox0.01h/{\rm Mpc}$. 

In comparison, for the projected clustering measured by $C_\ell$, lensing becomes non-negligible across all angular multipoles $\ell$ at redshift $z\gtrapprox0.5$ \cite{11ChallinorLPS,11Bonvin_GR}. In particular, for high $\ell$ at small angular scales, lensing dominates over the Newtonian RSD contribution and becomes the leading correction in addition to the density term $b_1D_{\rm m}$, so lensing needs to be included for modeling the galaxy clustering and its cross-correlation with cosmic shear and CMB lensing even in the Stage-III photometric surveys. We will discuss more on the behavior of the lensing term under the SFB basis in \cref{sec:lensing}. 

Due to the use of different approximations of the galaxy number count under different settings, different groups of terms in \cref{eq:GR} are referred to as the GR effects in different literature. In this work, we refer to all terms beyond the density and the standard RSD terms (the first line of \cref{eq:GR}) as the GR effects. Since the SFB power spectrum is of most interest for a wide-field spectroscopic survey, we benchmark against \cref{eq:spec}, the standard approximation used in spectroscopic surveys.

\subsection{Primordial Non-Gaussianities}\label{sec:PNG}

We have so far reviewed the relativistic galaxy number count for a Universe with Gaussian \HL{primordial} fluctuations. However, different models of the early universe such as various inflationary models could lead to a certain level of primordial non-Gaussianities (PNG). In particular, at large scales, the LSS is primarily sensitive to the so-called PNG of the local type, which includes a local, quadratic correction \HL{to} the primordial gravitational potential $\phi_{\rm p}$ parametrized by $f_{\rm NL}$ \cite{01Komatsu_fNL}:
\begin{align}
\phi_{\rm p}(\bx)=\phi_{\rm G}(\bx)+f_{\rm NL}\left[\phi_{\rm G}(\bx)^2-\langle\phi_{\rm G}(\bx)^2\rangle\right]\label{eq:fNL}.
\end{align}

Here we introduce the additional variable $\phi\equiv-\Phi$ ($\Phi$ is the Bardeen potential defined in \cref{eq:CNG}) to be consistent with the sign convention of $f_{\rm NL}$ used in literature such that a positive $\fNL$ enhances the power in the galaxy power spectrum. In single-field, slow-roll, models of inflation, local PNG are, for all practical purposes, negligible \cite{03Maldacena_single_field,04Creminelli_single_field}, while multi-field models of inflation generically predict large local PNG values $f_{\rm NL}\sim \mathcal{O}(1)$ \cite{10Chen_PNG_review,12Senatore_EFT,14fnl_target}, so a local PNG provides a smoking gun signature of those models. Measuring the PNG is an important goal of next-generation LSS surveys such as SPHEREx. 

Besides generating a primordial matter bispectrum, the local PNG also introduces a very distinct scale-dependent bias on the largest scales for the biased tracers \cite{07Dalal_PNG,08Slosar_fnl}, and this characteristic scale-dependent bias is observable from the two-point clustering. Under the local PNG, the bias relation between the biased tracers and the underlying matter density field at the linear order now becomes:
\begin{align}
D_{\rm g} = b_1D_{\rm m}+f_{\rm NL}b_{\phi}\phi \label{eq:bias-PNG},
\end{align}
where $b_{\phi}$ describes the response of galaxy formation to the large-scale potential $\phi$ at the time of structure formation. Even though the bias relation is usually discussed and formulated in the Newtonian limit, the only requirement to make the bias expansion consistent in the relativistic formalism at linear order is that the time slicing (constant-time hypersurface) chosen to perform the bias expansion should correspond to a constant proper time of comoving observers, which coincides with the synchronous-comoving gauge\footnote{The synchronous gauge is defined such that the perturbations of the time-time part and the space-time part of the metric tensor are equal to zeros. In this gauge, the time coordinate corresponds to the proper time of comoving observers at fixed spatial coordinates. In the absence of pressure, the synchronous gauge coincides with the comoving gauge at the linear order~\cite{06Hwang_sync_comv_2nd}, so one can use the two names interchangeably.}~\cite{09Wands,11Baldauf_GR_bias,18Desjacques_Galaxy_Bias}. Therefore, both the galaxy density $D_{\rm g}$ and the matter density $D_{\rm m}$ in \cref{eq:bias-PNG} shall be interpreted as quantities in the synchronous-comoving gauge, which is consistent with our notation for the relativistic number count in \cref{eq:GR}.

Assuming halo abundance models with a universal mass function (UMF), the halo bias with respect to the gravitational potential satisfies~\cite{08Slosar_fnl,15Tellarini_bias} 
\begin{align}
    b_{\phi}(b_1)=2\delta_{\rm c}(b_1-1),
    \label{eq:bphi_universal}
\end{align}
where $\delta_{\rm c}=1.686$ is the threshold for spherical collapse\footnote{$\delta_{\rm c}=1.686$ holds for the spherical collapse in both the Newtonian limit and the synchronous-comoving gauge in GR \cite{09Wands,11Baldauf_GR_bias}.}. However, the UMF assumption breaks down for galaxy samples selected from real surveys, and \cref{eq:bphi_universal} no longer holds for these realistic tracers. In this case, one could then measure $b_\phi$ from simulations~\cite{20Barreira_PNG_bias,23Lazeyras_bphi} or real data~\cite{23Sullivan_bphi} and marginalize over the priors on $b_\phi$ for constraining $f_{\rm NL}$~\cite{20Barreira_bphi_impact_constraint,22Barreira_bphi_BOSS}.

The Bardeen potential and the matter density in the comoving gauge are related by the relativistic Poisson equation 
\cite{82Bardeen,95Ma}:
\begin{equation}
\nabla^2\Phi(\bx,z) =\frac{3}{2}\mH^2(z)\Omega_{\rm m}(z)D_{\rm m}(\bx,z),
\label{eq:rel-Poisson}
\end{equation}
which is the same form as the Newtonian Poisson equation. In GR, the Poisson equation relates the Bardeen potential $\Phi$ (coincides with the potential in the conformal Newtonian Gauge) and the gauge-invariant density contrast $D_{\rm m}$ (coincides with the matter density in the comoving gauge). In the Fourier space, at the matter-dominated epoch $z_{\rm md}$, \cref{eq:rel-Poisson} becomes
\begin{align}
\Phi(k,z_{\rm md})&=-\frac{3}{2}\frac{\mH^2(z_{\rm md})}{k^2}\Omega_{M}(z_{\rm md})D_{\rm m}(k,z_{\rm md}).
\end{align}
We can rewrite the above equation in terms of the matter density at the source redshift $z$ (the redshift of the observed galaxy):

\begin{align}
    \Phi(k,z_{\rm md})&=-\frac{D_{\rm m}(k,z)}{\alpha(k,z)}\label{eq:primordial_potential},
\end{align}
where
\begin{align}
\alpha(k,z)&=\frac{2k^2\tilde{D}(z)(1+z_{\rm md})T(k)}{3\mH^2(z_{\rm md})\Omega_{\rm m}(z_{\rm md})}=\frac{2k^2\tilde{D}(z)T(k)}{3H_0^2\Omega_{\rm m0}}.
\end{align}
Here $\tilde{D}(z)$ is the growth factor normalized to the scale factor $(1+z)^{-1}$ at the matter-dominated epoch,  $T(k)$ is the matter transfer function, and $\Omega_{{\rm m}0}$ and $H_0$ are the matter density and the Hubble parameter today. Combining \cref{eq:bias-PNG} and \cref{eq:primordial_potential}, one obtains
\begin{align}
D_{\rm g} = \left(b_1+\frac{f_{\rm NL}b_{\phi}}{\alpha(k,z)}\right)D_{\rm m}\label{eq:scale-dependent-fnl},
\end{align}
which recovers the scale-dependent bias relation of local PNG initially derived in the Newtonian limit \cite{07Dalal_PNG,08Slosar_fnl}. 

Therefore, the commonly adopted \cref{eq:scale-dependent-fnl} remains true under GR at the linear order\footnote{However, one should take more care and re-examine whether the common modeling approach of galaxy bias and local PNG still holds in the relativistic context at the second order, where galaxy bias and gauge issues become more nuanced \cite{15Bertacca_second_order_bias,19Umeh_second_GR_bias,19Umeh_fnl_GR_bias,23Yoo}, and it might have implications for constraining local PNG through measuring the higher-order statistics of galaxy clustering near the horizon scale.}, and the power spectrum of galaxies still acquires the characteristic $k^{-2}$ signature on large scales. As $k\to 0$, the local PNG term diverges, which is physical and consistent with GR: the divergence is only acquired through the relativistic Poisson equation, while the galaxy bias $b_{\phi}$ remains constant, which suggests that the physical effect of long-wavelength fluctuations on the
local overdensity does not blow up. Strictly speaking, the term \HL{``scale-dependent bias"} ubiquitously used to describe the local PNG effect is a misnomer, as discussed in Ref.~\cite{22Barreira_bphi_BOSS}, since neither $b_1$ nor $b_\phi$ have any scale-dependence themselves, and it is the relativistic Poisson equation that depends on scales.

This $k^{-2}$ signature on large scales for the local PNG effect is shared by the gravitational potential term in \cref{eq:GR}. The PNG and GP terms are both directly proportional to the Bardeen potential $\Phi$, which scales with $k^{-2}$ through the relativistic Poisson equation derived from the linearized Einstein equations. The difference is that the PNG term depends on the potential at the matter-dominated epoch, while the GP term depends on the potential at the source redshift. From a heuristic argument, we can see that the GP term in GR effects can be degenerate with the scale-dependent $f_{\rm NL}$ signal, since they essentially have the same form proportional to the gravitational potential.

\section{SFB Basis and Power Spectrum}\label{sec:SFB}
In this section, we focus on the theoretical formalism behind the SFB power spectrum. We first briefly review the spherical Fourier-Bessel decomposition in both continuous and discrete cases following Ref.~\cite{21Gebhardt_SuperFab}. We then emphasize the benefit of using the discrete SFB basis over the continuous one. 

\subsection{SFB Decomposition}\label{sec:SFB-decomp}
The continuous SFB basis is composed of eigenfunctions of the Laplacian in spherical coordinates, namely the spherical Bessel functions of first kind $j_\ell(kx)$ and the spherical harmonics $Y^*_{\ell m}(\hat{\bn})$. The SFB decomposition of the density field $\delta(\bx)$ and the inverse transformation are:
\begin{align}
\delta_{\ell m}(k) &=
\int_{\bx}
j_\ell(kx)Y^*_{\ell m}(\hat{\bn})
\delta(\bx),
\label{eq:sfb_k-to-x}\\
\delta(\bx) &= \frac{2}{\pi}\int_{0}^{\infty}dk\,k^2\sum_{\ell m} j_\ell(kx)\,Y_{\ell m}(\hat{\bn})
\delta_{\ell m}(k)\,,
\label{eq:sfb_x-to-k}
\end{align}
where $\bx=x(z)\hat{\bn}$ is the position vector in the configuration space. The factor $2k^2/\pi$ can be split between \cref{eq:sfb_x-to-k,eq:sfb_k-to-x} as pleased. The continuous SFB power spectrum is defined as:
\begin{align}
   \langle\delta_{\ell_1 m_1}(k_1) \delta_{\ell_2 m_2}^{*}(k_2)\rangle = C_{\ell_1}(k_1,k_2)\delta_{\ell_1\ell_2}^{K}\delta_{m_1m_2}^{K}
   \label{eq:SFB-PS}\,,
\end{align}
where $\delta^{K}$ is the Dirac-delta symbol. 

In a homogeneous and isotropic Universe of infinite size with no LOS effects, $C_{\ell}(k_1,k_2)\sim P(k_1)\delta^{D}(k_1-k_2)/k_1^2$ for all $\ell$, where the SFB power spectrum is diagonal and proportional to the 3D power spectrum $P(k)$~\cite{13Yoo_GR_SFB,21Gebhardt_SuperFab}. This translational invariance is broken in the presence of redshift evolution and redshift space distortions, while the rotational invariance is still preserved, leading to the angular mode $\ell$ and two wavenumbers $k_1,k_2$ in \cref{eq:SFB-PS} as a complete decomposition of the two-point statistics for the density field. The two wavenumbers are obtained via the spherical Bessel transform, which is the Fourier transform in the radial direction, so $k_1$ and $k_2$ are the canonical Fourier modes.

Now \cref{eq:sfb_k-to-x} is the SFB decomposition of an infinite universe, where the Fourier modes are continuous, and for any $k$, $j_\ell(kx)Y^*_{\ell m}(\hat{\bn})$ is the eigenfunction of the Laplacian without any boundary conditions. Due to the finite speed of light, we only observe a finite universe, where a continuous decomposition of a field with only finite volume becomes redundant. The finite volume leads to discretization, which is similar to the discrete Fourier transform where the finiteness in the time domain leads to discrete modes in the frequency domain. In the presence of boundary conditions at the radial direction for some finite comoving distance range $x_{\rm min}\leq x \leq x_{\rm max}$, the orthogonality relation leads to discrete radial modes $k_{nl}$, and the radial eigenfunction of Laplacian becomes \cite{19Samushia_SFB,21Gebhardt_SuperFab}:
\begin{align}
g_{n\ell}(x) = c_{n\ell} \, j_\ell(k_{n\ell}x) + d_{n\ell} \, y_\ell(k_{n\ell}x),
\label{eq:gnl_basis}
\end{align}
which are linear combinations of spherical Bessel functions of the first and second kind, chosen to satisfy the orthonormality relation
\begin{align}
\label{eq:gnl_orthonormality}
\int_{x_{\rm min}}^{x_{\rm max}}dx\,x^2\,g_{n\ell}(x)\,g_{n'\ell}(x)
&=
\delta^K_{nn'}\,.
\end{align}

Here the index $n$ denotes the index for wavenumber $k_{n\ell}$ at each angular multipole $\ell$, and $c_{n\ell}$ and
$d_{n\ell}$ are constants that are determined by the boundary conditions. With the eigenvalue $-k_{n\ell}^2$, the function $g_{n\ell}(x)$ now serves as the eigenfunction to the radial component of the Laplacian under a chosen boundary condition. It is common to use either the potential or velocity boundary conditions, which are given in Appendix D of Refs.~\cite{21Gebhardt_SuperFab} and \cite{23Gebhard_SFB_eBOSS} respectively. In practice, we can compute $g_{n\ell}(x)$ of these two boundary conditions with the public code \href{https://github.com/hsgg/SphericalFourierBesselDecompositions.jl}{\texttt{SphericalFourierBesselDecompositions.jl}}. Generally, the differences between the two boundary conditions are small, and we choose the potential boundary condition throughout this work. 

The discrete SFB decomposition of the density field $\delta(\bx)$ and the inverse transformation then become 
\begin{align}
\delta_{n \ell m}
&= \int_{\bx}\,g_{n\ell}(x)\,Y^*_{\ell m}(\hat{\bn})\delta(\bx)\,,
\label{eq:sfb_discrete_fourier_pair_b}\\
\delta(\bx)
&= \sum_{n\ell m} g_{n\ell}(x)\,Y_{\ell m}(\hat{\bn})\,\delta_{n\ell m}\,,
\label{eq:sfb_discrete_fourier_pair_a}
\end{align} 
where the integral of \cref{eq:sfb_discrete_fourier_pair_b} goes over the finite volume within $x_{\rm min}\leq x \leq x_{\rm max}$. One can compare the above equations with the continuous SFB transform in \cref{eq:sfb_k-to-x,eq:sfb_x-to-k} to see the integral with respect to $k$ is reduced to a sum over $n$ due to the discretization, which is analogous to the summation replacing the integral in the discrete Fourier transform. We can then form the discrete SFB power spectrum through:
\begin{align}
   \langle\delta_{n_1 \ell_1 m_1}\delta_{n_2 \ell_2 m_2}^{*}\rangle = C_{\ell_1 n_1n_2}\delta_{\ell_1\ell_2}^{K}\delta_{m_1m_2}^{K}
   \label{eq:SFB-discrete-PS}\,.
\end{align}

\subsection{Advantage of the Discrete Basis}\label{sec:discrete-benefit}
In the canonical Fourier space, the continuous Fourier transform is used for analytical calculations where integration over an infinite range is both natural and possible, while data estimations and numerical computations always rely on the discrete Fourier transform due to a finite spatial or time domain. Similarly, the continuous SFB (CSFB) transform will be useful for analytical predictions to develop intuition, while one should apply the discrete SFB (DSFB) transform for extracting information on cosmological data and numerically computing the signal. 

In this work, we will calculate GR effects in the DSFB power spectrum. To fully justify this choice, we will explain the practical advantages of the DSFB basis in this section, besides drawing an analogy to the discrete Fourier transform. The DSFB basis (extending from $x_{\rm min}$ to $x_{\rm max}$ with radial functions $g_{n\ell}(x)$) is pioneered by Ref.~\cite{19Samushia_SFB} and then used for estimators and theory calculations in Refs.~\cite{21Gebhardt_SuperFab,22Khek_SFB_fast,23Gebhard_SFB_eBOSS}. Our advocacy for the discrete basis is therefore not new, but we here provide a more detailed and explicit summary of its benefits and practical applications.

\begin{figure}[htbp!]
\centerline{\includegraphics[width=0.92\hsize]{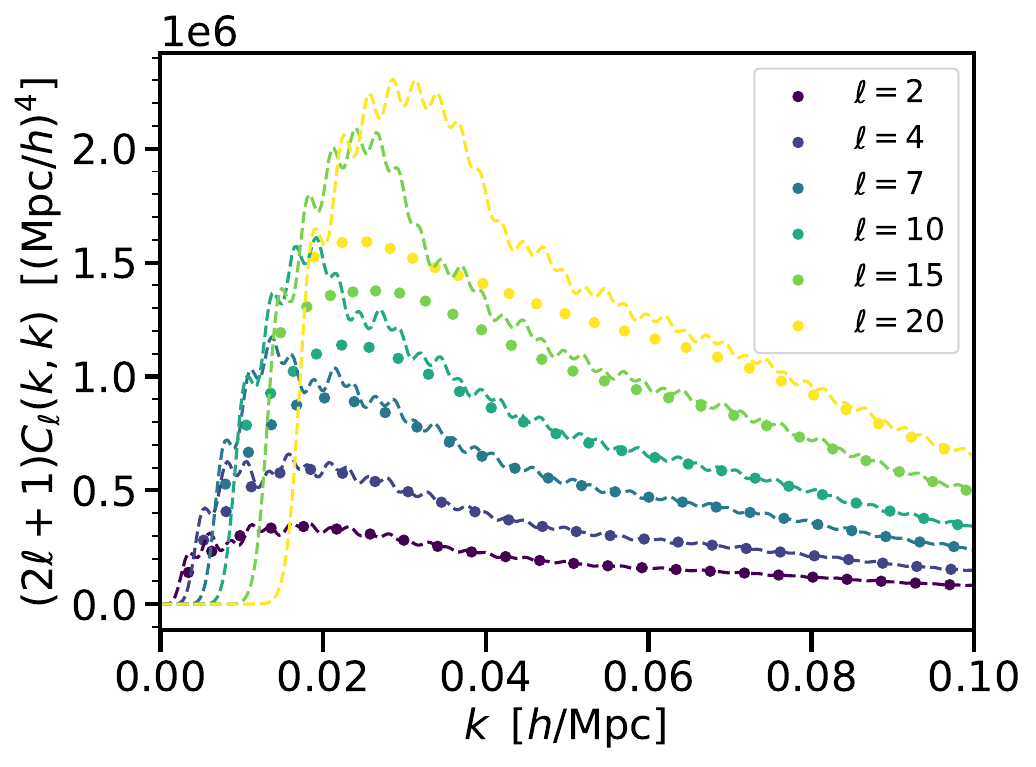}}
\caption{Comparing the continuous and discrete SFB power spectra for a uniform radial selection with a finite redshift range from $z_{\rm min}=0.2$ to $z_{\rm max}=0.5$. The points show the DSFB power spectrum, while the dashed lines show the CSFB. We assume the linear Newtonian RSD modeling in \cref{eq:spec} and $b_1=1.5$. Since it is impossible to properly normalize the CSFB basis in a finite volume, the CSFB and DSFB PS will not match in general. To roughly match the two spectra for comparison, we apply a single normalization factor to the continuous SFB PS at all $\ell$ and $k$. An additional $2(\ell+1)$ factor is applied on both the discrete and the continuous SFB PS to distinguish the different angular multipoles in the plot. We see that the DSFB spectrum is smooth and well-behaved since DSFB is a complete basis over the finite volume, while the CSFB spectrum suffers from oscillations due to the non-smoothness of the top-hat window in a finite redshift range.}
\label{fig:cSFB_compare_dSFB}
\end{figure}

In the case that the density volume reaches $x_{\rm min}=0$, the DSFB radial function is proportional to the continuous one at the discretized wavenumbers, that is $g_{n\ell}(x)=c_{n\ell}j_{\ell}(k_{n\ell}x)$, so the two power spectra will coincide up to some normalization. However, most tracers in realistic galaxy surveys start at some minimum redshift such that $x_{\rm min}(z_{\rm min})\neq 0$, in which case the DSFB radial basis function will become a generic linear combination of spherical Bessel functions of both first and second kinds as given in \cref{eq:gnl_basis}, so the DSFB and CSFB spectra will become different. Assuming a finite redshift coverage from $z=0.2$ to $0.5$, we compare the CSFB and DSFB power spectra in \cref{fig:cSFB_compare_dSFB}, which are computed following \cref{sec:compute-numerics}. Since the CSFB is not an orthonormal basis in a finite volume, it is hard to properly normalize the CSFB PS to exactly match the DSFB PS. We apply a single normalization factor to the continuous SFB PS at all $\ell$ and $k$ in \cref{fig:cSFB_compare_dSFB} to roughly match the two spectra for comparison\footnote{Due to the use of a single normalization for the CSFB at all $\ell$ and $k$, the DSFB and CSFB PS are not exactly matched in \cref{fig:cSFB_compare_dSFB}, they become more different at higher multipoles, which is not surprising since the radial basis functions in DSFB and CSFB are different. The main point of \cref{fig:cSFB_compare_dSFB} is to show the stability of the DSFB over CSFB.}. We see that the discrete spectrum is smooth and well-behaved, while the CSFB spectrum suffers from oscillations due to the abrupt drop of galaxy density at the boundaries of the finite redshift range. Therefore, the DSFB PS is a more stable quantity when the redshift range starts at non-zero redshift.

Besides its numerical stability and aesthetic appeal, the DSFB basis has the theoretical advantage of being a complete basis of the density field over a finite volume. \HL{Discretization} imposes a discrete sampling of the SFB PS that can be efficiently represented in a matrix form $C_{\ell n_1n_2}\equiv C_{\ell}(k_{n_1\ell},k_{n_2\ell})$. For a Gaussian universe, the DSFB PS will contain all the information with a finite number of data points, serving as a lossless compression of the density field. For the CSFB, one has to choose a fine but arbitrary resolution in the Fourier $k$ modes during numerical computation, and results still need to be binned for practical purposes. In comparison, the DSFB naturally sets the $k$ resolution and avoids the arbitrary binning step. As seen in \cref{fig:cSFB_compare_dSFB}, there are only about 20 $k_{n\ell}$ modes in the linear regime ($k\lessapprox0.1\,h/{\rm Mpc}$) for a full-sky volume from $z=0.2$ to $0.5$. Due to its efficient and complete decomposition of the field, the DSFB PS can be mapped to other two-point statistics, and the discrete basis simplifies and stabilizes the numerical implementation of these mappings as we will discuss in \cref{sec:SFB-map}.

Furthermore, discretization explicits how the Fourier $k_{n\ell}$ modes depend on and vary with respect to angular multipoles $\ell$. For each angular mode $\ell$, DSFB naturally gives the lowest Fourier mode $k_{0\ell}$, representing the largest scale accessible at the given angular scale $\ell$. In general, only the largest angular scales at multipoles $\ell\lessapprox 10$ contribute to the lowest Fourier modes at $k\lessapprox0.01\,h/{\rm Mpc}$, demonstrated in  \cref{fig:cSFB_compare_dSFB}. Such explicit relationships between the angular and Fourier modes can help us better understand and control the various angular and redshift systematics in real surveys. 

Besides the Fourier mode $k$ and angular multipole $\ell$, the DSFB basis also gives access to an additional index variable $n$, which does not exist for the CSFB. \HL{This additional index $n$ properly discretizes the Fourier modes, and} in \cref{sec:lensing} we will see that most of the lensing effect is contained in the $\HL{n=0}$ modes\footnote{\HL{Here we label the discrete Fourier modes (as defined in Eq.~\ref{eq:gnl_basis}) starting from $n=0$ instead of starting from $n=1$ as done in the previous works \cite{21Gebhardt_SuperFab,23Gebhard_SFB_eBOSS}. The largest Fourier mode accessible at each angular multipole now corresponds to $n=0$, which better characterizes the property that these SFB basis functions possess minimal radial oscillations.}}, so these modes will be useful for measuring the lensing signal. We will also see that the index $n$ in the DSFB basis is somewhat analogous to the variable $\mu\equiv k_{||}/k$, parametrizing the component parallel to the LOS, often used in the plane-parallel limit. We expect $n$ to provide a useful handle for isolating different combinations of angular and radial modes in LSS surveys.

It is also more natural to use the DSFB basis to build estimators that can be applied to data, and a pseudo-$C_{\ell}$ estimator for the DSFB power spectrum has been fully developed in Ref.~\cite{21Gebhardt_SuperFab}. Therefore, the theoretical computation should match the DSFB estimator for validation and inference. For all these reasons, we recommend the use of the DSFB basis over the continuous one for numerical computation. In the rest of the paper, we simply use \HL{``SFB" }to refer to the DSFB basis, and we will explicitly mention \HL{``continuous"} when we refer to CSFB.

\section{Computation of the SFB Power Spectrum}\label{sec:SFB_compute}
We now describe our method for computing the SFB power spectrum. We first summarize all the analytical formulas used for computing the SFB PS and GR effects in \cref{sec:compute-analytic}, and we will develop some intuition based on these formulas. We next review the numerical approach proposed in Sec.~IV of Ref.~\cite{23Gebhard_SFB_eBOSS} for computing the SFB PS under the Newtonian RSD in \cref{sec:summary_numerics}. We then adapt this approach to compute the GR effects in \cref{sec:GR_SFB}. 

\subsection{Analytical Formalism}\label{sec:compute-analytic}

According to Refs.~\cite{22Khek_SFB_fast,23Gebhard_SFB_eBOSS}, the SFB power spectrum given in \cref{eq:SFB-discrete-PS} can be computed using the following integral:
\begin{align}
    C_{\ell n_1n_2}^{\rm R}&=\int_{0}^{\infty}dq\, \mathcal{W}_{n_1\ell}^{\rm R}(q) \mathcal{W}_{n_2\ell}^{\rm R}(q) P_{\rm m,0}(q)\label{eq:dSFB_compute}\,,
\end{align}
where $P_{\rm m,0}(q)$ is the present matter power spectrum under the comoving gauge, which can be computed from the linear Boltzmann codes. The SFB kernel $\mathcal{W}_{n\ell}(q)$ is given by
\begin{align}
\mathcal{W}_{n\ell}^{\rm R}(q)\equiv\sqrt{\frac{2}{\pi}}q\int_{x_{\rm min}}^{x_{\rm max}}dx\,x^2g_{n\ell}(x)R(x)\Delta_{\ell}(x,q)\label{eq:Wnlq_kernel}\,.
\end{align}
We have added a radial selection function $R(x)$ to the SFB PS, and we use the subscript R to indicate this case. Here $\Delta_{\ell}(x,q)$ is the angular kernel that can contain various physical effects including the Newtonian RSD, GR effects~\cite{11Bonvin_GR,13DiDio_classgal}, redshift error modeling, and Fingers-of-god effects~\cite{22Khek_SFB_fast}. We will review the derivation of the above two formulas in \cref{sec:SFB-kernel}.

To compute the GR effects in the SFB PS using \cref{eq:dSFB_compute,eq:Wnlq_kernel}, we summarize the angular kernels $\Delta_{\ell}(x, q)$ corresponding to all the linear-order GR effects (ignoring the observer's terms) following Ref.~\cite{11Bonvin_GR,13DiDio_classgal}:
\begin{widetext}
\begin{subequations}
\begin{align}
&\Delta_{\ell}(x, q)^{\rm DRSD}=b_1(x)D_{\rm m}(x, q)j_{\ell}(q x)+\frac{q}{\mathcal{H}(x)}v(x, q) j_{\ell}''(q x)\label{eq:D_R}\\
&\Delta_{\ell}(x, q)^{\rm Doppler}=\mathcal{A}_1(x) v(x, q)j_{\ell}'(qx) \label{eq:D_D} \\
&\Delta_{\ell}(x, q)^{\rm NIP}=\Bigg[\left(\mathcal{A}_1(x)+1\right)\Psi(x, q)-(2-5s(x)) \Phi(x, q)+\frac{1}{\mathcal{H}(x)}\dot{\Phi}(x, q)\left(b_{\rm e}(x)-3\right)\mathcal{H}(x)V(x, q)\Bigg]j_{\ell}(q x)\label{eq:D_G}\\
&\Delta_{\ell}(x, q)^{\rm Lensing}=\ell(\ell+1)\frac{2-5 s(x)}{2}\int_{0}^{x}dr\,\frac{x-r}{x r}(\Phi(r, q)+\Psi(r, q))j_{\ell}(q r) \label{eq:D_L}\\
&\Delta_{\ell}(x, q)^{\rm Shapiro}=\frac{2-5 s(x)}{x}\int_{0}^{x}dr\, (\Phi(r, q)+\Psi(r, q)) j_{\ell}(q r)\label{eq:D_S}\\
&\Delta_{\ell}(x, q)^{\rm ISW}=\mathcal{A}_1(x)\int_{0}^{x}dr\, \left(\dot{\Phi}(r, q)+\dot{\Psi}(r, q)\right)j_{\ell}(q r)\,.\label{eq:D_I}
\end{align}
\end{subequations}
\end{widetext}
Here $V(x(z),q)$ represents the transfer function for the velocity potential from the present-time matter perturbation $D_{{\rm m},0}(q)$; the same convention is followed for the transfer functions $\Phi(x,q)$ and $\Psi(x,q)$ for the two Bardeen potentials. These angular kernels have been given in Refs.~\cite{11ChallinorLPS,11Bonvin_GR,13DiDio_classgal} to compute the GR effects in the angular power spectrum, and they can be directly derived from the galaxy number count expression in \cref{eq:GR}. For completeness, we will review the derivation of the above, as well as the angular kernels for the observer's terms, in \cref{sec:GR_kernel}. 

For a Universe with local non-Gaussian primordial fluctuations, an additional scale-dependent component to the angular kernel is added:
\begin{align}
&\Delta_{\ell}(x, q)^{\rm PNG}= \frac{f_{\rm NL}b_{\phi}(x)}{\alpha(q,x)}D_{\rm m}(x,q)j_{\ell}(qx).
\label{eq:D_PNG}
\end{align}

The transfer functions for the velocity and gravitational potentials depend on the particular choice of cosmology, and in general they need to be calculated from the linear Boltzmann codes such as \texttt{CAMB} and \texttt{CLASS}. However, in the standard $\Lambda$CDM cosmology at the late time where one can ignore the anisotropic stress caused by radiation and neutrinos, the Friedmann equation and the linearized Einstein equations\footnote{The Bardeen potential transfer function in \cref{eq:approx-Phi} is derived from the relativistic Poisson equation in \cref{eq:rel-Poisson}, and it was already used in \cref{sec:PNG} to explain the scale-dependent bias generated from the local PNG.} give the following~\cite{03Dodelson,22CatorinaGR-P,23WAGR}:
\begin{subequations}
 \begin{align}
&D_{\rm m}(x,q)=D(x)\label{eq:approx-D}\\
&v(x,q)=qV(x,q)=-f(x)\frac{\mH(x)}{q}D(x)\label{eq:approx-v}\\
&\Phi(x,q)=\Psi(x,q)=-\frac{3}{2}\frac{\mH^2(x)}{q^2}\Omega_{\rm m}(x)D(x)\label{eq:approx-Phi}\\
&\dot{\Phi}(x,q)=\dot{\Psi}(x,q)\nonumber\\
&\qquad\quad=-\frac{3}{2}\frac{\mH^3(x)}{q^2}\Omega_{\rm m}(x)(f(x)-1)D(x)\label{eq:approx-dot-Phi}\\
&\frac{\dot{\mH}}{\mathcal{H}^2}=1-\frac{3}{2}\Omega_{\rm m}(x)\,,
\end{align}
\end{subequations}
where $D(x)$ is the linear growth factor normalized at the current time and $f(x)$ is the linear growth rate. In the $\Lambda$CDM model, the transfer functions are separable in terms of the conformal time indicated by the comoving distance $x(z)$ and the Fourier mode $k$. In this work, we will assume the $\Lambda$CDM model and use the above transfer functions to calculate the GR effects. 

The above expressions for transfer functions directly indicate how individual GR effects scale with the Fourier mode as one moves closer to the horizon. The Doppler term is proportional to the velocity scalar $v$, which scales as $k^{-1}$ according to \cref{eq:approx-v}. The GP term, including the NIP, Shapiro, and ISW effects, is proportional to the Bardeen potentials and their time derivative, which scale as $k^{-2}$ seen from \cref{eq:approx-Phi,eq:approx-dot-Phi}. Proportional to the angular Laplacian of the Bardeen potentials, the lensing term will also scale as $k^{-2}$. 

We can also predict the angular dependence of individual GR terms through the analytical formulas summarized in this section. Since the transfer functions do not contain any angular dependence, we see that there is no explicit dependence on $\ell$ besides the spherical Bessel function $j_{\ell}(qx)$ in the angular kernels for most GR terms. The exception is the lensing effect, which explicitly contains an $\ell(\ell+1)$ factor in its angular kernel given by \cref{eq:D_L}. We can therefore naively predict the lensing effect to scale as approximately $\ell^{2}$. These analytical predictions of dependence on $\ell$ and $k$ will be confirmed by the numerical results in \cref{sec:lk-dependence}. 

\subsection{Numerical Computation}\label{sec:compute-numerics}
Our overall numerical strategy is to first integrate along $x$ to compute the SFB kernel in \cref{eq:Wnlq_kernel} and then perform the integration along $q$ to obtain the SFB PS following \cref{eq:dSFB_compute}. This is also the integration order followed by previous SFB literature~\cite{13Yoo_GR_SFB,14Nicola_SFB,15Lanusse_SFB,21Zhang_SFB_TSH,22Khek_SFB_fast,23Gebhard_SFB_eBOSS}, since \cref{eq:dSFB_compute,eq:Wnlq_kernel} reduce the 3-dimensional (3D) integrals contained in the SFB PS to an effective 2D integral.

\subsubsection{Summary of Iso-qr Integration}\label{sec:summary_numerics}

The main difficulty in computing the SFB power spectrum lies in the numerical evaluation of the SFB kernel $\mathcal{W}_{n\ell}(q)$. In the absence of GR effects with only the Newtonian RSD, Ref.~\cite{23Gebhard_SFB_eBOSS} has proposed an efficient procedure to compute the SFB kernel such that the evaluation of the SFB power spectrum only takes seconds on a single CPU, enabling inference of cosmological parameters through MCMC with the SFB PS. 

The key insight realized by Ref.~\cite{23Gebhard_SFB_eBOSS} is that the bulk of the computation for the SFB kernel in \cref{eq:Wnlq_kernel} is spent on evaluating the spherical Bessel functions $j_{\ell}(qx)$ inside the angular kernel $\Delta_{\ell}(x,q)$. We can put the formula for the SFB kernel into the general form of
\begin{align}
    \int_{x_{\rm min}}^{x_{\rm max}}dx\,f(x,q)j_{\ell}(qx)\,.\label{eq:qr_integration}
\end{align}
In the SFB kernel, the radial basis function $g_{n\ell}(x)$ will be precomputed, while the angular kernels $\Delta_{\ell}(x,q)$ for the DRSD term, as shown in \cref{eq:D_R}, only involves simple algebraic operations except the spherical Bessel functions $j_{\ell}(qx)$, so the evaluation of $f(x,q)$ is fast, while computing $j_{\ell}(qx)$ is slow.

Ref.~\cite{23Gebhard_SFB_eBOSS} further notices that the spherical Bessel functions only depend on the variable combination $qx$, not on $q$ and $x$ separately. Thus, we can choose discretizations for $q$ and $x$ such that in q-x space the “iso-qx” lines go precisely through the grid points of $q$ and $x$. Using the “iso-qx” lines transforms the problem of calculating the spherical Bessel function from a 2D to a 1D problem: we can efficiently precompute and cache the values of spherical Bessel functions only on the “iso-qx” lines and then perform the 1D integration in \cref{eq:qr_integration}. We refer to this procedure as the Iso-qr integration. Ref.~\cite{23Gebhard_SFB_eBOSS} has already given an outline of this algorithm, and we will supplement the details in \cref{sec:qr_integration} and extend the algorithm to perform the Iso-qr integration over multiple subintervals in \cref{sec:interval_integration}. 

\subsubsection{Adaptation for GR effects} \label{sec:GR_SFB}

The Iso-qr integration scheme has allowed the efficient computation of the SFB power spectrum under Newtonian RSD in Ref.~\cite{23Gebhard_SFB_eBOSS}. In this work, we will use the same scheme extended for all the linear-order GR effects. The effects from \cref{eq:D_R} to \cref{eq:D_I} can be classified into two classes: the non-integrated terms and the integrated terms, the latter containing an extra integration along the LOS. The calculation of the SFB kernel $\mathcal{W}_{n\ell}(q)$ for the non-integrated terms can be directly achieved using the Iso-qr integration scheme without any modification: the standard Newtonian RSD term has already been implemented in Ref.~\cite{23Gebhard_SFB_eBOSS}, and the Doppler and NIP terms are simple extensions.

For the integrated terms including the lensing, Shapiro, and ISW terms (\cref{eq:D_I,eq:D_L,eq:D_S}), the SFB kernels contain an additional integration along the LOS in the angular kernel $\Delta_{\ell}(x,q)$. Fortunately, evaluating these integrated terms only requires small modifications to the Iso-qr integration method, which we will explain below.

To simplify the implementation, we first observe that the lensing term in \cref{eq:D_L} can be written as a sum:
\begin{align}
&\Delta_{\ell}(x, q)^{\rm Lensing}=\left(\frac{2-5 s(x)}{2}\right)\ell(\ell+1)\nonumber\\
&\qquad\quad\Bigg[\int_{0}^{x}dr\, \frac{1}{r}(\Phi(r, q)+\Psi(r, q))j_{\ell}(q r)\nonumber\\
&\qquad\quad-\frac{1}{x}\int_{0}^{x}dr\, (\Phi(r, q)+\Psi(r, q))j_{\ell}(q r)\Bigg],
\end{align}
where the two integrands inside the integral along the LOS do not depend on the source redshift $x(z)$. The angular kernels of all three integrated terms can then be cast to the following form:
\begin{align}
    &\Delta_{\ell}(x,q)^{\rm I}=K(x)\int_{0}^{x}dr\,g(r,q)j_{\ell}(qr)\nonumber\\
    &=K(x)\left[\int_{0}^{x_{\rm min}}dr\,g(r,q)j_{\ell}(qr)+\int_{x_{\rm min}}^{x}dr\,g(r,q)j_{\ell}(qr)\right]\nonumber\\
    &=K(x)G(q)+\Delta_{\ell}(r=x_{\rm min}\to x,q)^{\rm I}\label{eq:split},
\end{align}
where we have split the LOS integral into two parts. The integration from $0$ to $x_{\rm min}$ does not depend on the source position $x$, so this part of the integral can be precomputed using the Iso-qr integration scheme over $0$ to $x_{\rm min}$ outside the SFB kernel evaluation loop and stored as \HL{a} 1-D array $G(q)$.

We next examine the integration step from $x_{\rm min}$ to $x$. The SFB kernel in \cref{eq:Wnlq_kernel} has the following form for the integrated terms (from $x_{\rm min}$ to $x$):
\begin{align}
\mathcal{W}_{n_1\ell}^{\rm I}(q)&\sim\int_{x_{\rm min}}^{x_{\rm max}}dx\,M(x)\int_{x_{\rm min}}^{x}dr\,g(r,q)j_{\ell}(qr)\label{eq:WI-integrate}\,.
\end{align}
We can choose the outer integral and the inner integral in \cref{eq:WI-integrate} to share the same discretization scheme $\Delta x_{i}=\Delta r_{i}$ over the range $x_{\rm min}$ to $x_{\rm max}$ for the trapezoidal rule such that at a particular $q$: 
\begin{align}
\mathcal{W}_{n_1\ell}^{\rm I}(q)\sim \sum_{i=0}^{N}\Delta x_{i}\,M(x_i)\sum_{j=0}^{i}\Delta r_{j}\,g(r_j,q)j_{\ell}(qr_j)\label{eq:integrated_tem_discretized}\,.
\end{align}
Since the inner integrand in \cref{eq:WI-integrate} does not depend on $x$, the integration only needs to be performed for the outer integral, while we increment the inner integrand over the same interval as the following:
\begin{align}
R_{i+1}&\leftarrow R_{i}+\Delta x_{i}\,g(x_i,q)j_{\ell}(qx_i)\,,\label{eq:increment}\\
\mathcal{W}_{n_1\ell}^{\rm I}&\sim \sum_{i=0}^{N}\Delta x_{i}\,M(x_i)R_i\,.
\end{align}
The above equations effectively rewrite \cref{eq:integrated_tem_discretized} into a weighted sum of a cumsum (cumulative sum), so the integrand only needs to be evaluated $N$ times where $N$ is the number of discretized bins from $x_{\rm min}$to $x_{\rm max}$. This reduces the 2D integral in \cref{eq:WI-integrate} to a 1-D summation, and one can obtain the discretization $\Delta x_{i}$ following the Iso-qr integration scheme to evaluate $j_{\ell}(qx)$. 

Recall that the SFB kernel for the non-integrated terms has the form:
\begin{align}
     \mathcal{W}_{n_1\ell}^{\rm NI}(q)&\sim\int_{x_{\rm min}}^{x_{\rm max}}dx\,f(x,q)j_{\ell}(qx)\,\label{eq:WNI-integrate}.
\end{align}
We can use the same \HL{``iso-qx"} line from the Iso-qr integration scheme to evaluate the spherical Bessel functions for both non-integrated and integrated terms, since they cover the same range from $x_{\rm min}$ to $x_{\rm max}$. In summary, the computational complexity for the integrated terms remains the same order as the non-integrated terms. The integrated terms simply require an additional precomputation step to integrate from $0$ to $x_{\rm min}$ following \cref{eq:split}, and then increment the integrand following \cref{eq:increment} to integrate from $x_{\rm min}$ to $x_{\rm max}$.

\begin{figure}[tbp]
\centerline{\includegraphics[width=0.95\hsize]{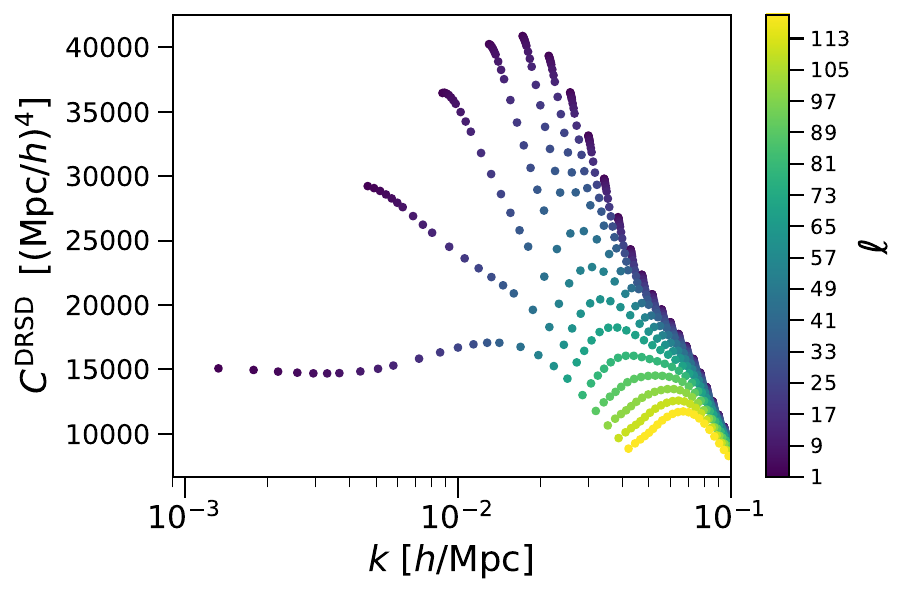}}
\caption{The DRSD (density plus RSD) term in the diagonal part of the SFB power spectrum $C_{\ell nn}\equiv C_{\ell}(k_{n\ell},k_{n\ell})$ for a redshift range from $z=1.0$ to $1.5$. The colormap shows the corresponding angular multipoles $\ell$. We only plot a subset of all the $\ell$ multipoles for visualization. }
\label{fig:DRSD_10_15}
\end{figure}

\begin{figure*}[tbp]
\centerline{\includegraphics[width=0.75\textwidth]{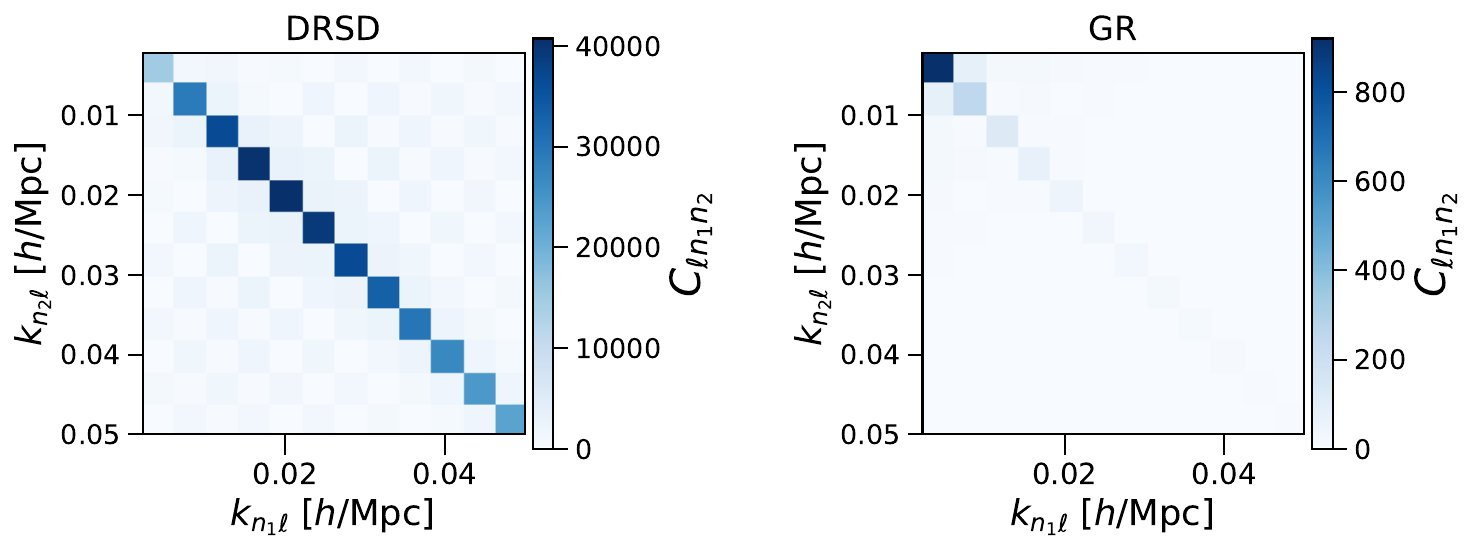}}
\caption{The full SFB power spectrum $C_{\ell n_1n_2}\equiv C_{\ell}(k_{n_1\ell},k_{n_2\ell})$ (including both the diagonal and off-diagonal parts) in $[{\rm Mpc}/h]^4$ at $\ell=3$ for a redshift range from $z=1.0$ to $1.5$. The left panel shows the DRSD term, while the right panel shows the total linear-order GR correction. With a uniform radial window, the SFB PS is mostly diagonal.}
\label{fig:off-diagonal}
\end{figure*}

\begin{figure*}[tbp]
\centerline{\includegraphics[width=0.73\textwidth]{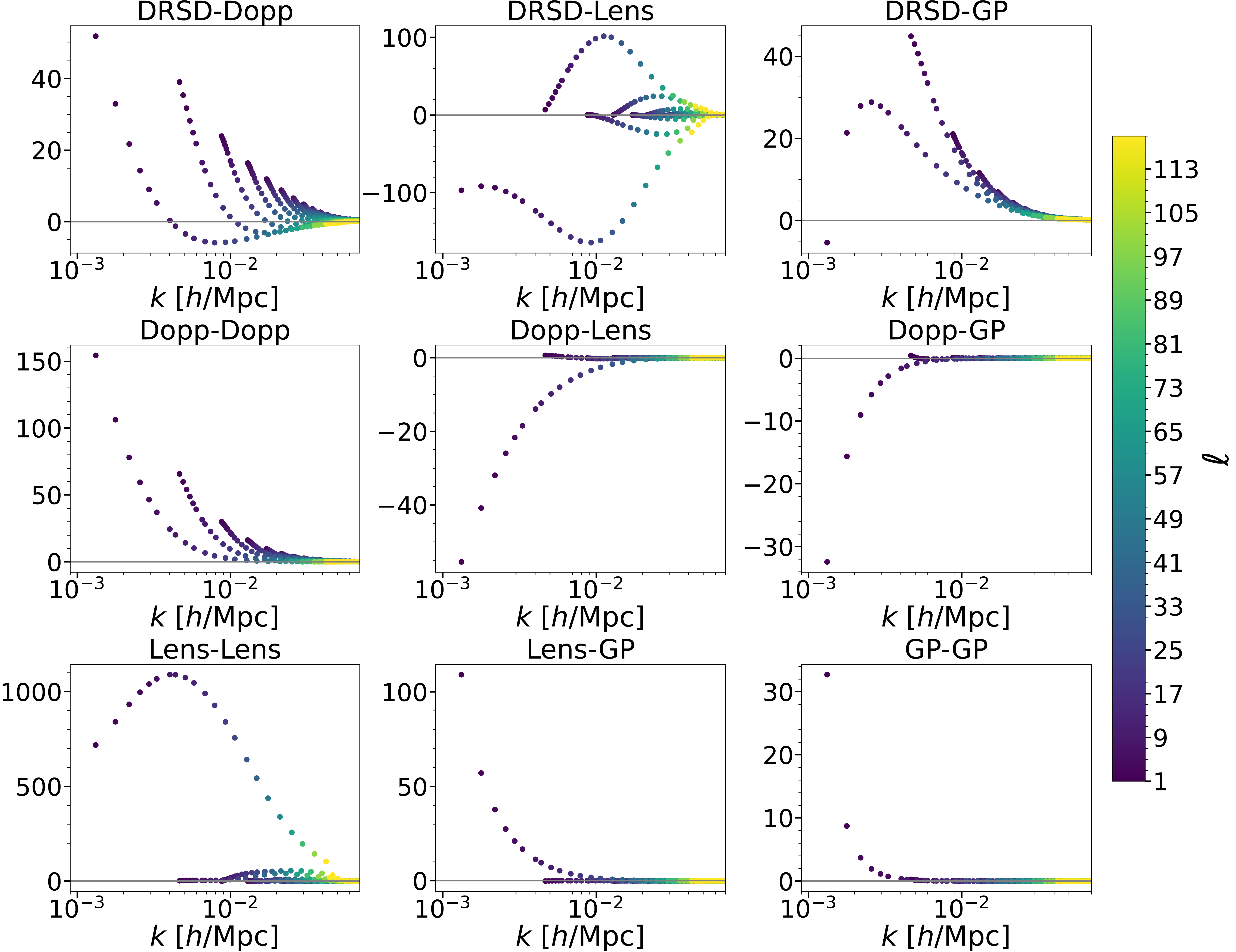}}
\caption{The GR terms in the diagonal part of the SFB power spectrum $C_{\ell nn}$ in $[{\rm Mpc}/h]^4$ for a redshift range from $z=1.0$ to $1.5$. The colormap shows the corresponding angular multipoles $\ell$. The thin grey lines indicate the $0$ value for the SFB PS. We only choose a subset of all the $\ell$ multipoles for visualization.}
\label{fig:GR_10_15}
\end{figure*}

\section{Results}\label{sec:result}
Here we present our results of GR effects for SFB multipoles $\ell\geq 1$. We will discuss the SFB monopole in more detail in \cref{sec:divergence}. Our numerical implementation has been validated by comparing the angular power spectra transformed from our SFB calculations using \cref{eq:TSH-SFB} and results directly obtained from \texttt{CLASS}, which we show in \cref{sec:validation}.

For all results in this section, we set the linear bias $b_1=1.5$, and we consider a vanishing magnification bias $s=0$ and evolution bias $b_{\rm e}=0$. Different values for the evolution and magnification biases will only change the amplitude of the GR corrections, and we do not expect our qualitative conclusions to change under other realistic values for these two biases. We assume the universal bias relation \cref{eq:bphi_universal} for any calculations related to the local PNG. We use the notation $C^{AB}$  to indicate the cross-correlation between the effects $A$ and $B$. The diagonal part of the SFB power spectrum $C_{\ell}(k_{n\ell},k_{n\ell})$ is the same regardless of the ordering of effects entering the cross-correlation, that is $C^{AB}$ and $C^{BA}$ have the same diagonal SFB PS. For reference, we show an example of the diagonal part of the SFB PS for the standard density plus the Newtonian RSD (DRSD) term in \cref{fig:DRSD_10_15}. 

With the numerical formalism described in \cref{sec:compute-numerics}, we can compute the full SFB power spectrum including both the diagonal ($n_1=n_2$) and the off-diagonal ($n_1\neq n_2$) terms. For a uniform radial function in the comoving space as we considered in this work, the off-diagonal terms are substantially smaller compared to the diagonal part of the spectrum for both the standard DRSD term and the total linear-order GR correction, as shown in \cref{fig:off-diagonal}. Therefore, we will focus our discussion on the diagonal terms in this work. With a non-uniform radial window as the case for real surveys, the diagonal part of the spectrum will then be mixed to the off-diagonal part, leading to more substantial contribution to the off-diagonal components~\cite{22Khek_SFB_fast,23Gebhard_SFB_eBOSS}.

\subsection{$\ell$ and $k$ Dependence}\label{sec:lk-dependence}

We first show the SFB PS for the auto-correlation and cross-correlation among all GR effects in \cref{fig:GR_10_15} for a redshift range from $z=1.0$ to $1.5$ as an example. In general, GR corrections are only important at low Fourier modes corresponding to large scales, and all GR effects significantly decrease for higher $k$ values. The panels of \cref{fig:GR_10_15} demonstrate the rich and different patterns of the individual GR terms in the SFB $\ell$-$k$ space. We next separately examine the $\ell$ and $k$ dependence of the GR terms in detail.

\begin{figure}[tbp]
\centerline{\includegraphics[width=0.92\hsize]{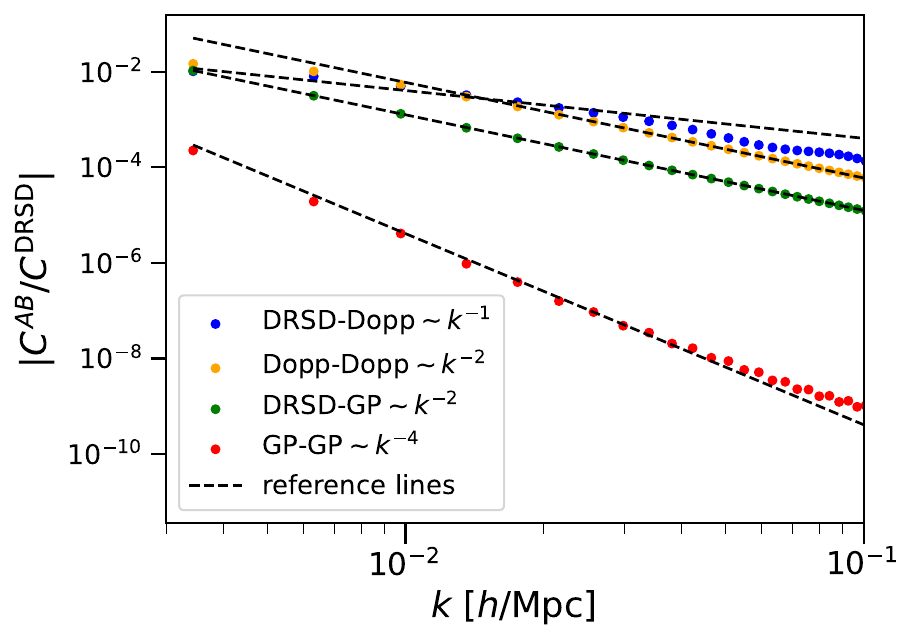}}
\caption{The $k$ dependence at $\ell=2$ for the diagonal part of the SFB PS $C_{\ell nn}$ for a redshift range from $z=0.2$ to $0.5$. We plot the absolute values of the ratios between the individual GR terms and the DRSD term. The dots represent the SFB PS, while the black dashed lines indicate the corresponding scaling behavior with respect to $k$. }
\label{fig:k_scale_Doppler_GP}
\end{figure}

\begin{figure}[tbp]
\centerline{\includegraphics[width=0.92\hsize]{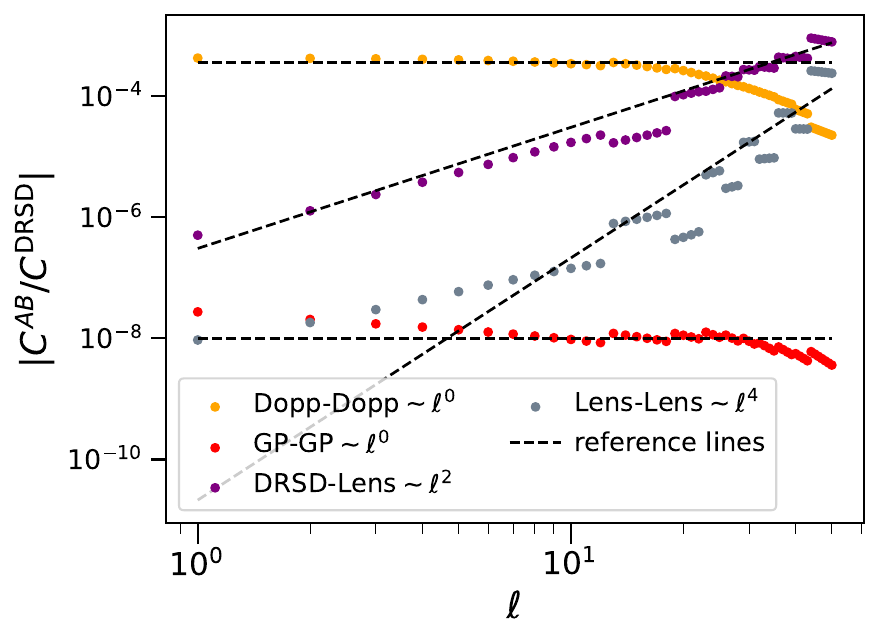}}
\caption{The $\ell$ dependence at $k \approx 0.04\,h/{\rm Mpc}$ for the diagonal part of the SFB PS $C_{\ell nn}$ for a redshift range from $z=0.2$ to $0.5$. We plot the absolute values of the ratios between the individual GR terms and the DRSD term. The dots represent the SFB PS, while the black dashed lines indicate the corresponding scaling behavior with respect to $\ell$. Since the set of discrete Fourier modes $k_{n\ell}$ are different for each angular multipole, it is impossible to exactly fix the Fourier mode while varying $\ell$. To examine the angular dependence in the right subplot, we identify the $k_{n\ell}$ mode closest to $0.04\,h/{\rm Mpc}$ for each multipole $\ell$ as an approximation for fixing $k$. The inability of exactly fixing $k$ leads to some spurious oscillation patterns around the reference lines.}
\label{fig:l_scale_Doppler_GP}
\end{figure}

Fixing the angular multipole $\ell$, we show the scaling behaviors of the GR terms with the Fourier $k$ mode in \cref{fig:k_scale_Doppler_GP}, and they conform to our expectation based on the transfer functions given in \cref{eq:approx-D,eq:approx-v,eq:approx-Phi,eq:approx-dot-Phi}. Proportional to the velocity scalar $v$, the DRSD-Doppler term scales roughly as $k^{-1}$ at large scales. In comparison, the DRSD-GP term scales almost exactly as $k^{-2}$ predicted by the transfer function for the potential $\Phi$. For the auto-correlation, the Doppler-Doppler and GP-GP terms scale as $k^{-2}$ and $k^{-4}$ respectively, in line with the k-scaling behavior for $v^2$ and $\Phi^2$. The lensing-lensing term also scales approximately as $k^{-4}$, seen in the right panel of \cref{fig:lensing_fixed_n} below, since it is also proportional to $\Phi^2$. Therefore, the Doppler effect has a shallower $k$ dependence compared to other GR effects, and it is the only GR effect proportional to the velocity field instead of the gravitational potential. Other cross-correlation GR terms such as Doppler-lensing, Doppler-GP, and lensing-GP terms shown in \cref{fig:GR_10_15} also exhibit power-law scaling relation with $k$.

We next examine the angular dependence of the GR terms. The Doppler and GP terms do not exhibit any noticeable angular dependence as shown in \cref{fig:l_scale_Doppler_GP}. In contrast, the DRSD-lensing and lensing-lensing terms increase roughly as $\ell^2$ and $\ell^4$ as one considers larger angular multipoles $\ell$ corresponding to smaller angular scales. These scaling relationships can be fully explained by the $\ell(\ell+1)$ factor in the lensing angular kernel \cref{eq:D_L}. The angular Laplacian in the lensing convergence (\cref{eq:kappa}) imprints this unique angular dependence, setting lensing terms apart from all other linear-order GR terms. We will discuss more about lensing in \cref{sec:lensing}.

\subsection{Evolution across Redshift}

\begin{figure*}[tbp]
\centerline{\includegraphics[width=0.82\hsize]{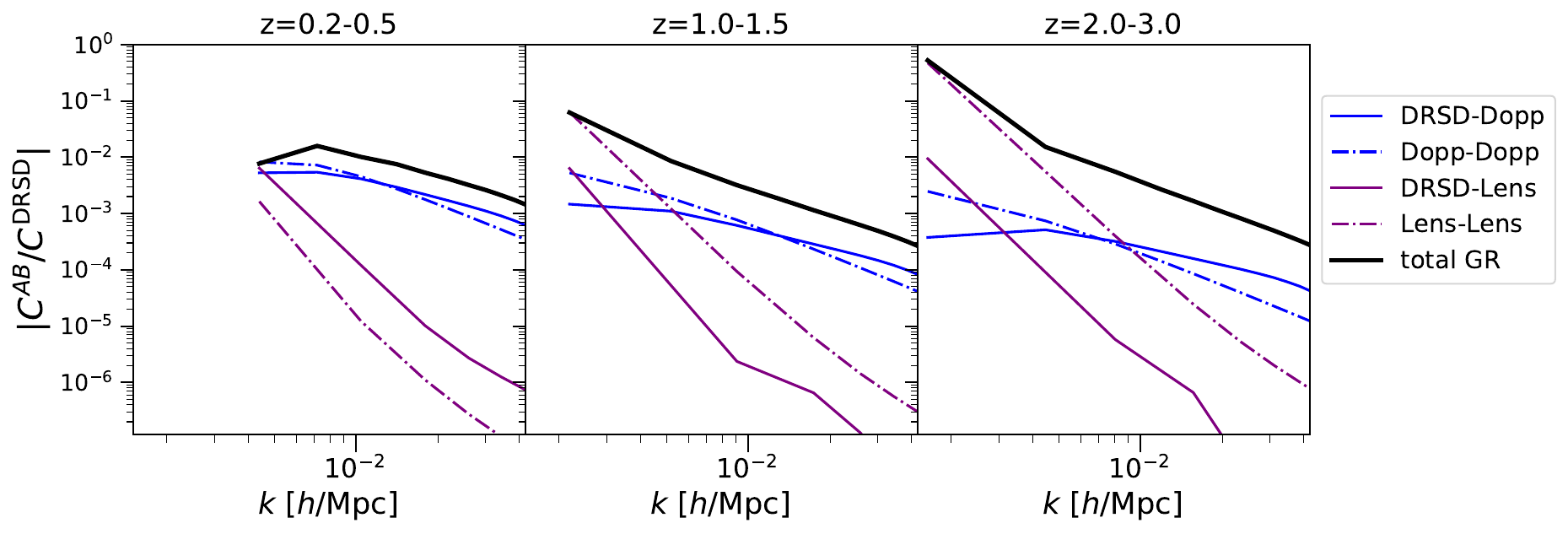}}
\caption{The importance of total and individual GR effects including Doppler and lensing terms in three redshift bins at $\ell=3$. Here we use lines instead of points to plot the discrete SFB PS to make the plots more readable. For the DRSD-lensing and lensing-lensing terms, we only plot the $k_{n\ell}$ modes where $n$ is odd to avoid showing the oscillations due to the parity of $n$, which can be seen in the \cref{fig:lensing_fixed_n} below.}
\label{fig:GR_term_evolve}
\end{figure*}

\begin{figure*}[tbp]
\centerline{\includegraphics[width=0.82\textwidth]{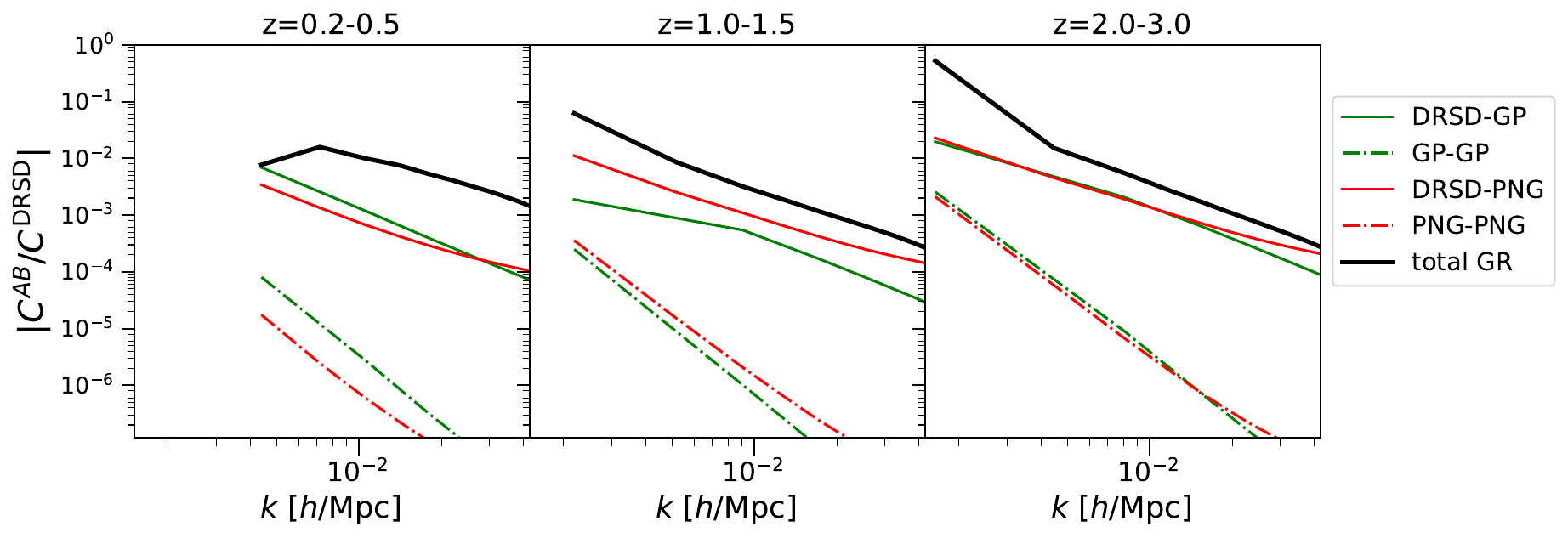}}
\caption{The importance of PNG and GP terms in three redshift bins at $\ell=3$. The PNG terms are computed assuming $f_{\rm NL}=1$.}
\label{fig:PNG_term_evolve}
\end{figure*}

We next discuss how GR terms evolve across the redshift. We show the relative significance of individual and total GR effects, normalized by the SFB PS under Newtonian DRSD, in three different redshift ranges in \cref{fig:GR_term_evolve,fig:PNG_term_evolve} at $\ell=3$. The total GR effect increases as one considers higher redshift: the GR correction is only around the percent level at the lowest redshift bin $z=0.2-0.5$, increasing to several percent for $z=1.0-1.5$, and reaching tens of percent for the high redshift bin $z=2.0-3.0$. This drastic increase in the significance of GR effects underlines the importance of modeling these terms for galaxy clustering at high redshift, which can be probed by the current Stage-IV and the proposed Stage-V surveys~\cite{19MSE,22Schelegel_megamapper,22Schlegel_S5}.

The increase of total GR effects at the high redshift is driven by the integrated terms including the lensing, Shapiro, and ISW effects, which accumulate from the observer to the source along the LOS. \cref{fig:GR_term_evolve,fig:PNG_term_evolve} show the increasing importance of lensing terms and GP terms respectively at higher redshift. Lensing terms become non-negligible for $z=1.0-1.5$, reaching a few percent at the lowest Fourier modes, and it drastically increases to tens of percent for $z=2.0-3.0$. In both redshift bins, the auto-correlation lensing-lensing term dominates over the cross-correlation DRSD-lensing term by about an order of magnitude. In comparison, the cross-correlation DRSD-GP term dominates over the auto-correlation GP-GP term as seen in \cref{fig:PNG_term_evolve}. The RSD-GP term remains at the sub-percent level for the two lower redshift bins, and it increases above the percent level for the highest redshift bin $z=2.0-3.0$.

While the lensing and GP effect increase at higher redshift, the Doppler effect gradually decreases in significance as seen in \cref{fig:GR_term_evolve}. The Doppler effect, including the cross-correlation DRSD-Doppler term and the auto-correlation Doppler-Doppler term, is the dominant GR correction at the lowest redshift bin $z=0.2-0.5$, reaching the percent level, but it becomes negligible compared to other GR effects at the highest redshift bin $z=2.0-3.0$. In summary, the Doppler effect is important at low redshift, while lensing is dominant at high redshift. The redshift evolution pattern we observe in the SFB space is qualitatively in agreement with previous literature such as Refs.~\cite{11ChallinorLPS,11Bonvin_GR,15Jeong_GR} and \cite{22CatorinaGR-P} that examine the individual GR effects in TSH and PSM respectively.

\begin{figure*}[tbp]
\centerline{\includegraphics[width=0.65\textwidth]{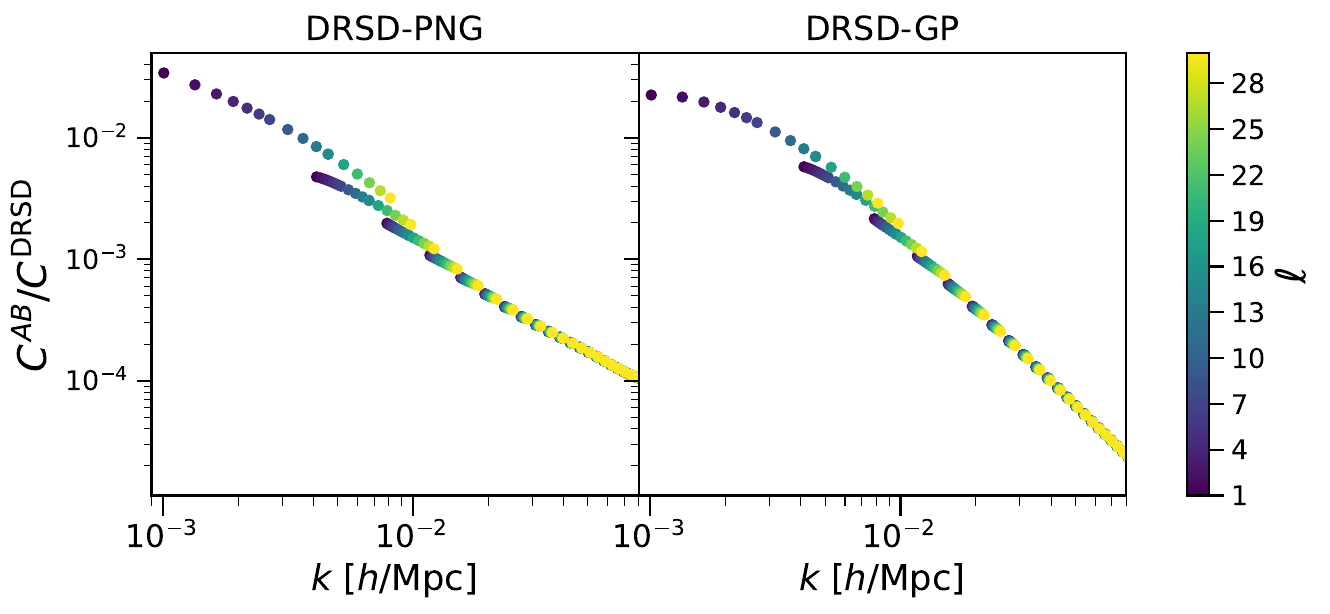}}
\caption{Comparison between the DRSD-PNG and DRSD-GP terms, normalized by the Newtonian DRSD term, in $C_{\ell nn}$ for $z=2.0$ to $3.0$. The DRSD-PNG term is calculated with $f_{\rm NL}=1$. We only show the SFB modes with low $\ell$ values for clarity, and the decreasing trend continues for higher $\ell$ multipoles.}
\label{fig:PNG_GP_compare}
\end{figure*}

\subsection{Degeneracy with PNG}

During our review of the local PNG in \cref{sec:PNG}, we discussed the possible degeneracy between the GP and PNG terms, since both effects are directly proportional to the gravitational potential $\Phi$. We now compare our numerical results for the DRSD-GP term and the DRSD-PNG term, which is evaluated at $f_{\rm NL}=1$, in the SFB space in \cref{fig:PNG_GP_compare}. We see that these two effects have almost the same dependence on $\ell$ and $k$, implying the degeneracy between the two signals. However, these two effects evolve slightly differently across the redshift, since the PNG effect depends on the potential during the matter-dominated era, while the GP effect contains the NIP term (depending on $\Phi$ at the source), the Shapiro term (depending on $\Phi$ integrated along the LOS), and the ISW term (depending on $\dot{\Phi}$ integrated along the LOS). In \cref{fig:PNG_term_evolve}, we see that the GP effect mimics a local PNG signal of unity for the redshift bin $z=2.0-3.0$, and it becomes slightly larger and smaller than $f_{\rm NL}\sim 1$ for $z=0.2-0.5$ and $1.0-1.5$ respectively. In general, the GP effect mimics $f_{\rm NL}\sim \mathcal{O}(1)$, so it will be important to model the GP effect for surveys aiming to constraint $\sigma(f_{\rm NL})\sim 1$ so as to avoid any bias in the measurement. 

Even though the GP effect is degenerate with the local PNG, the Doppler and lensing effects do not exhibit the same degeneracy, since the Doppler effect has a distinct $k$-dependence, while the lensing effect has a unique $\ell$-dependence as previously discussed. Therefore, the total GR effects are not completely degenerate with $f_{\rm NL}$ in the SFB space. However, to achieve consistency in modeling, GR effects should be considered for surveys reaching $\sigma(f_{\rm NL})$ to a few, since the total GR effect is noticeably larger than the local PNG effect at $f_{\rm NL}=1$ across the redshift as illustrated in \cref{fig:PNG_term_evolve}.

\subsection{Behavior of Lensing}\label{sec:lensing}
\begin{figure*}[tbp]
\centerline{\includegraphics[width=0.7\textwidth]{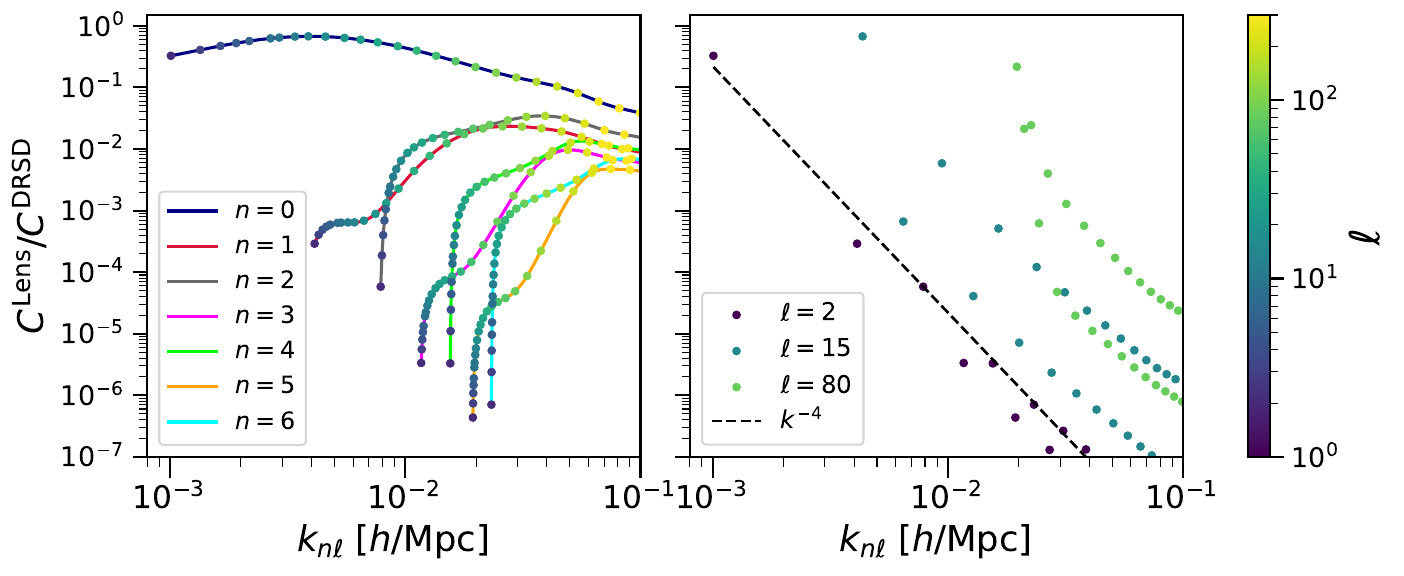}}
\caption{The behavior of the lensing-lensing term
for $C_{\ell nn}$, normalized by the SFB PS under the Newtonian DRSD term, for a redshift range from $z=2.0$ to $3.0$. The dots show the values of the SFB PS, with the angular modes indicated by the colormap. The left panel emphasizes the dependence of the lensing effect on $n$, and we use color dashed lines to connect the Fourier modes $k_{n\ell}$ with the same $n$. The dots only show a subset of all the $\ell$ multipoles at each $n$ for visualization. In the right panel, we plot the lensing effect at several particular angular multipoles $\ell$ to show the $k$-dependence.}
\label{fig:lensing_fixed_n}
\end{figure*}

We next focus on the behavior of the lensing-lensing term, since it is the most important correction at $z\gtrapprox 1$ as demonstrated in \cref{fig:GR_10_15,fig:GR_term_evolve}, dominating over other GR corrections by at an order of magnitude. The lower left panel of \cref{fig:GR_10_15} singles out one prominent branch of the SFB PS compared to other SFB modes. Further examination in the left panel of \cref{fig:lensing_fixed_n} shows that this dominant branch in the lensing-lensing term corresponds to the SFB $k_{n\ell}$ modes with varying $\ell$ at the constant $\HL{n=0}$. These \HL{$k_{0\ell}$} modes correspond to the largest scale (the lowest Fourier mode) accessible at each angular multipole. 

The left panel of \cref{fig:lensing_fixed_n} shows the prominent dependence of the lensing term on $n$. For a fixed $n$, the lensing effect remains relatively stable for smaller angular scales. On the iso-$n$ contour of an approximately horizontal direction, smaller Fourier scales (higher $k$ values) correspond to smaller angular scales (higher $\ell$ values), and the $\ell(\ell+1)$ factor imprinted by the angular Laplacian compensates the $k^{-2}$ scaling behavior in the potential $\Phi$ and $\Psi$, resulting in a relatively stable lensing correction even for smaller angular scales.

Fixing the Fourier mode $k$, we see that the lensing correction drastically decreases for higher $n$ values as seen in the left panel of \cref{fig:lensing_fixed_n}. This is because a higher $n$ mode on the iso-$k$ contour corresponds to a lower $\ell$ value (larger angular scales), which dramatically suppresses the lensing-lensing correction due to its approximate $\ell^4$ dependence. This explains the dominance of the $\HL{n=0}$ branch, since it has the highest $\ell$ mode for a given $k$.

In the plane-parallel limit, the transverse Fourier mode can be approximated by $k_{\perp}\approx\ell/r_{\rm eff}$ for a given angular multipole $\ell$, where $r_{\rm eff}$ is the effective comoving distance for a relatively narrow redshift bin, and the parallel Fourier mode can be given by $k_{||}^2=k_{n\ell}^2-k_{\perp}^2$. Therefore, higher angular multipoles correspond to higher $k_{\perp}$ values and smaller transverse scales. On the iso-$k$ contour (the vertical directions in \cref{fig:lensing_fixed_n}), smaller $n$ values correspond to higher $\ell$ values and lower $k_{||}$ values, which represent larger radial scales. Therefore, the $\HL{n=0}$ branch represents the largest radial scales accessible at each $\ell$ and the smallest angular scales (\HL{the purely angular scales}) accessible at each $k$. In the plane-parallel language, $\HL{n=0}$ can then be approximately characterized as $\mu\equiv k_{||}/k\approx 0$, where lensing is the most dominant.

Interestingly, there seems to be some parity dependence on $n$ for the lensing correction. Examining the left panel of \cref{fig:lensing_fixed_n}, we see that the branches with the same parity of $n$ (except the dominant $\HL{n=0}$ branch) demonstrate similar patterns. The lensing correction only strictly decreases for increasing $n$ values of the same parity, while the neighboring branches, such as \HL{$n=1$ and $2$} branches can intersect each other. This parity dependence can also be seen in the iso-$\ell$ contour plotted in the right subplot of \cref{fig:lensing_fixed_n}, where there are two distinct lines decreasing with higher $k$ for a fixed $\ell$.

The lensing pattern we see in the SFB space is consistent with previous works that examine the lensing effect in PSM and TSH. Refs.~\cite{22CatorinaGR-P,23WAGR} shows the lensing PSM $P_{L}(k)$ drastically decreases for higher $k$ values, while the lensing correction roughly remains stable in significance as $\ell$ increases as seen in Refs.~\cite{11ChallinorLPS,11Bonvin_GR} for the TSH case. The angular power spectrum sums over all Fourier modes $n$ (as explicitly shown in \cref{eq:TSH-SFB}), so the lensing correction in the TSH will be dominated by the $\HL{n=0}$ branch. The relative stability of the $\HL{n=0}$ branch explains the stability of the lensing effect with respect to $\ell$ in the TSH. Therefore, the SFB PS demonstrates the same $\ell$ and $k$ dependence for lensing as in previous works. The clear separation of the radial and angular scales in the SFB basis allows us to examine both sets of dependence at the same time and fully captures the rich pattern of the lensing convergence.

Last, we highlight the exciting opportunity of measuring the lensing convergence present in the galaxy clustering at high redshift through the SFB basis. \cref{fig:lensing_fixed_n} shows the lensing correction reaches tens of percent of the Newtonian DRSD clustering in the \HL{$k_{0\ell}$} modes across a wide range of $\ell$ at the redshift bin $z=2.0-3.0$. These $\HL{n=0}$ modes capture most of the lensing effect compared to the higher $n$ modes. Since the lensing effect becomes comparable to the density plus Newtonian RSD term, it will become measurable at these $\HL{n=0}$ modes. Therefore, it is important to model the lensing effect for the future Stage-V galaxy surveys aiming at the high redshift Universe for both achieving precise cosmological constraints and measuring the lensing convergence.

\section{Conclusion and Discussion}\label{sec:conclusion}

As the current and upcoming LSS surveys probe galaxy clustering over increasingly large volumes, the spherical Fourier-Bessel basis provides a powerful tool to model the general relativistic and wide-angle effects 
that become important at large scales. In this paper, we presented the first computation of the linear-order relativistic effects in the discrete SFB power spectrum. Compared to the continuous SFB basis used in previous works~\cite{13Yoo_GR_SFB,24Semenzato_SFB_GR} for modeling GR effects, the discrete basis we adopt here provides a more stable and efficient decomposition of the galaxy density field and matches the implementation of estimators. Our main numerical approach is based on the Iso-qr integration scheme introduced in Ref.~\cite{23Gebhard_SFB_eBOSS}, and we have adapted the method for the integrated GR terms including the lensing, Shapiro, and ISW effects. Moreover, we have validated our implementation of GR effects by transforming our SFB PS results to the TSH (through the mapping in \cref{eq:TSH-SFB}), which agree well with outputs from \texttt{CLASS}.

The SFB basis has allowed us to examine both angular $\ell$ and Fourier mode $k$ dependence of the GR effects at the same time, and we find that the Doppler effect has a shallower $k$ dependence compared to other GR effects, while the lensing effect has a unique $\ell$ dependence due to the angular Laplacian in the lensing convergence. The effect of local primordial non-Gaussianities shares a similar pattern in the $\ell$-$k$ space with the gravitational potential term, suggesting degeneracy between the two and the need of modeling the GP term when $\sigma(f_{\rm NL})\sim 1$ is reached. The Doppler effect is the most important at the low redshift, while the lensing convergence dominates at the high redshift. Due to the large impacts of the lensing effect on the $k_{n\ell}$ modes with $\HL{n=0}$ (which correspond to the largest scales at each angular multipole) at the high redshift, upcoming LSS surveys can potentially measure the lensing convergence present in galaxy clustering by measuring these SFB modes. In general, GR effects become increasingly important at higher redshift, and these effects should be considered when we can constraint $f_{\rm NL}$ to a few.

We have so far considered all linear-order GR effects in the SFB space, except the observer's terms. Considering a full-sky survey with no angular mask, the observer's potential only impacts the SFB monopole, while the observer's velocity only impacts the SFB dipole (as shown in \cref{sec:observer}). For cosmological inference with the SFB PS, one can simply exclude the first two multipoles and use angular modes $\ell\geq 2$ similar to the CMB analysis~\cite{18Planck_Parameter}. Therefore, the observer's terms are of no concern for the main cosmological analysis of galaxy surveys. 

However, they are still interesting quantities for both theory and observation. The observer's potential is instrumental for the cancellation of infrared divergence in the GR effects at the SFB monopole, which we will discuss more in \cref{sec:divergence}. The presence of the observer's velocity in the dipole offers the opportunity to test the cosmological principle in the SFB basis. The observer's velocity, which is our kinematic velocity with respect to the CMB rest frame and also more commonly known as the kinematic dipole, has been measured from LSS tracers such as quasars \cite{24Mittal_quaia_dipole,24Abghari_dipole} or radio sources \cite{02Blake_dipole,24Cheng_radio_dipole} using the projected statistics purely in the angular space. These measurements from LSS are in various degrees of tension with respect to the CMB measurement~\cite{17Colin_dipole,22Secrest_dipole,23Wagenveld_dipole,2018HFIcalibration,2018LFIcalibration,2020PlanckNPIPE}. Compared to the projected statistics in the angular space, the SFB basis contains the full 3D information and offers a better handle on systematics, so it provides another promising avenue for measuring the kinematic dipole. We will discuss all these issues related to the observer's terms in depth in forthcoming works.

We emphasize that our calculation of the relativistic effect is performed under the strict assumption of the standard $\Lambda$CDM model under General Relativity. Ref.~\cite{16Munshi} has explored modified gravity models in the SFB basis, albeit with the Newtonian modeling of RSD. To extend our calculation of general relativistic SFB to non-$\Lambda$CDM models, one has to obtain the transfer functions $V(z,k)$, $\Psi(z,k)$ and $\Phi(z,k)$ from modified Boltzmann models such as \texttt{MGCAMB} \cite{19Zucca_MGCAMB}, \texttt{EFTCAMB} \cite{14Hu_EFTCAMB}, and \texttt{hiclass} \cite{17Zumalacarregui_HiClass}. All numerical techniques used in this work remain effective under relatively smooth velocity and potential transfer functions, and we did not assume the separability between the redshift evolution $z$ and the Fourier mode $k$ in these transfer functions. Therefore, to extend our calculation of the relativistic effects to non-$\Lambda$CDM models, we will only need to integrate our SFB implementation with a Boltzmann code. Such an implementation will allow a rigorous test of modified gravity or alternative dark energy theories at the perturbation level with galaxy clustering on the largest scales, similar to how the low-$\ell$ part of the CMB power spectra provide constraints on these non-$\Lambda$CDM models at the perturbation level~\cite{15Planck15_MGDE,18Planck_Parameter}.

Assuming the standard $\Lambda$CDM model under GR, our validated calculation of the linear-order relativistic correction in the SFB PS can be used for the cosmological analysis of large-scale clustering. However, further optimization and computation speedup will be required to use our calculation for MCMC inference. With the Iso-qr integration scheme, we can compute the SFB power spectrum including all GR effects in a couple of seconds with a single CPU, for a lower redshift bin $z\lesssim 0.5$ considering all the radial Fourier modes $k_{n\ell}\lesssim 0.1 h/{\rm Mpc}$. However, the computation can take tens of seconds when one considers higher redshift or smaller scales. As suggested in Ref.~\cite{21Zhang_SFB_TSH}, further computation speedup on evaluating the SFB PS might be achieved by using a FFTLog-based method similar to Refs.~\cite{17Assassi_TSH,18Gebhardt_FFTlog,18Schoneberg_cl_fftlog_gr,20Fang_FFTlog} for the angular power spectrum. Our Iso-qr integration scheme has the advantage of not requiring the potential transfer functions including $V(z,k)$ and $\Phi(z,k)$ to be separable, and it does not require apodization and zero-padding schemes employed in FFTLog. Our implementation can serve as the baseline for a more optimized implementation of the SFB PS in the future. 

Our calculation of GR effects in the SFB PS will also have implications for the modeling of GR effects in other two-point statistics, since the SFB basis is a complete decomposition of the Gaussian density field and can be transformed into other statistics (as shown in \cref{sec:SFB-map}). In particular, we will be able to use the SFB-to-PSM \HL{mapping} introduced and developed by Refs.~\cite{18CastorinaWA,24PSM_SFB,24gSFB} to calculate the GR effects present in the PSM, which has been the preferred statistics for Fourier-space clustering analysis due to the efficient implementation of its estimator, efficient representation of RSD effects, and well-developed non-linear scale modeling. We will also be able to compute the PSM covariance with GR effects using the mapping between the PSM covariance and SFB PS introduced in Ref.~\cite{24PSM_SFB}, which will then allow us to forecast the measurability of GR effects and assess their impact on the $f_{\rm NL}$ constraints from PSM for future surveys. Our theoretical modeling of linear-order GR effects in the SFB basis will play an important role to the analysis of the large-scale clustering to be measured in the upcoming LSS surveys.

Our work can be extended to the SFB bispectrum, of which the numerical evaluation is recently achieved in Ref.~\cite{23Benabou_SFB} under Newtonian RSD. Modeling GR effects for the SFB bispectrum is significantly more challenging than the power spectrum case due to the very large number of terms present in the second-order GR effects~\cite{14Dio_GNC_2nd,14Yoo_second_order_GR,17Nielsen_higher_GR,18Bertacca_GR_SFB_Bi,19Fuentes_second_GR} and the higher dimensions of the integrals, but they can be important for the upcoming LSS surveys such as SPHEREx aiming at achieving $\sigma(f_{\rm NL})\sim 0.5$ through the bispectrum analysis~\cite{23WAGR,24Addis_Bi_GR}. Our work lays out the foundation for future works investigating higher-order GR effects in the discrete SFB basis, which will help us to fully utilize the cosmological information and probe fundamental physics at the largest scales in the observable Universe.

\section*{Acknowledgements} 
We thank the SPHEREx cosmology team for useful discussions. We acknowledge support from the SPHEREx project under a contract from the NASA/GODDARD Space Flight Center to the California Institute of Technology. Part of this work was done at Jet Propulsion Laboratory, California Institute of Technology, under a contract with the National Aeronautics and Space Administration (80NM0018D0004).

\appendix
\section{Mapping with SFB}\label{sec:SFB-map}

Since the SFB basis is a complete decomposition of the density field, one can express any other two-point statistics in terms of the SFB power spectrum. We will first give the mappings between the angular power spectrum and the SFB PS in \cref{sec:TSH-SFB}. These mappings will be helpful for deriving the SFB kernels in \cref{sec:SFB-kernel} and validating our GR results in \cref{sec:validation}. The mapping from the SFB PS to TSH can be further extended to the projected angular power spectrum and the correlation function multipoles $\xi_{L}(s,x)$, which we give in \cref{sec:pTSH-SFB} and \cref{sec:CF-SFB} respectively. The mapping from the SFB power spectrum to the power spectrum multipoles (PSM) was first given in Ref.~\cite{18CastorinaWA,24PSM_SFB} for the continuous SFB basis and then in Ref.~\cite{24gSFB} for the discrete case. This section will complete the study of transforming the SFB PS to other common two-point statistics used in galaxy surveys.  

\subsection{TSH-SFB Mapping}\label{sec:TSH-SFB}
As seen in \cref{eq:sfb_k-to-x,eq:sfb_discrete_fourier_pair_b}, the SFB transform relies on the spherical harmonics to obtain the angular modes. Angular multipoles $\ell$ are typically encountered in the context of the angular power spectrum. In fact, the angular $\ell$ modes share the same physical interpretation for both TSH and SFB, which we will demonstrate by giving the mappings between TSH and SFB.

The spherical harmonic mode $\delta_{\ell m}(x(z))$ in the configuration space is defined as
\begin{align}
    \delta_{\ell m}(x)=\int_{\hat{\bn}}Y^*_{\ell m}(\hat{\bn})\delta(\bx).
    \label{eq:TSH-x}
\end{align}
These spherical harmonic modes form the angular power spectrum:
\begin{align}
    \langle\delta_{\ell_1 m_1}(x)\delta_{\ell_2 m_2}(x')\rangle=C_{\ell_1}(x,x')\delta^{K}_{\ell_1\ell_2}\delta^{K}_{m_1m_2},
    \label{eq:TSH}
\end{align}
and it is more conventional to directly write the angular power spectrum in terms of the redshift $C_{\ell}(z,z')$ instead of the comoving distance.

We now observe the relationship between the SFB mode $\delta_{n\ell m}$ and the TSH mode $\delta_{\ell m}(x)$. From the definition of the SFB mode in \cref{eq:sfb_discrete_fourier_pair_b}, we have
\begin{align}
    \delta_{n \ell m}
&= \int_{x_{\rm min}}^{x_{\rm max}}dx\,x^2g_{n\ell}(x)\int_{\hat{\bn}}Y^*_{\ell m}(\hat{\bn})\delta(\bx)\nonumber\\
&=\int_{x_{\rm min}}^{x_{\rm max}}dx\,x^2g_{n\ell}(x)\delta_{\ell m}(x).
\end{align}
Therefore, the SFB power spectrum can be written as a two-dimensional integral transform of the angular power spectrum:
\begin{align}
    C_{\ell n_1n_2}=&\int_{x_{\rm min}}^{x_{\rm max}}dx\,x^2g_{n_1\ell}(x)\nonumber\\
&\quad \int_{x_{\rm min}}^{x_{\rm max}}dx'\,x'^2g_{n_2\ell}(x')C_{\ell}(x,x')\,\label{eq:dSFB-TSH}
\end{align}
which maps the TSH to the SFB space. The SFB PS and TSH that entered the above equation share the same value of $\ell$, and both statistics reside in the same angular space.

The above \cref{eq:dSFB-TSH} assumes a uniform selection in the comoving space within the redshift range. To include a radial selection function $R(x)$ in the SFB power spectrum (we use the superscript R to indicate this case), we can generalize \cref{eq:dSFB-TSH} to obtain:
\begin{align}
    C_{\ell n_1n_2}^{\rm R}=&\int_{x_{\rm min}}^{x_{\rm max}}dx\,x^2g_{n_1\ell}(x)R(x)\nonumber\\
&\quad \int_{x_{\rm min}}^{x_{\rm max}}dx'\,x'^2g_{n_2\ell}(x')R(x')C_{\ell}(x,x')\,.\label{eq:dSFBR-TSH}
\end{align}
Our SFB calculation in \cref{sec:SFB_compute} is formulated for the above case with the presence of a radial selection function. 

Since the numerical computation of the angular power spectrum has been extensively developed in literature \cite{17AngPow,18Gebhardt_FFTlog,17Assassi_TSH,19CCL,20Fang_FFTlog,23N5K}, with the linear-order GR effects included in Refs.~\cite{11ChallinorLPS,13DiDio_classgal,18Schoneberg_cl_fftlog_gr}, \cref{eq:dSFBR-TSH} raises the interesting question of whether it is possible to calculate the SFB PS relying on existing implementations of the TSH instead of using the SFB kernel as we did in \cref{sec:SFB_compute}. The main challenge of directly using the TSH-to-SFB map in \cref{eq:dSFB-TSH} for numerical evaluations is the non-trivial 2D integration of the angular power spectrum, and the total integration involved in evaluating the SFB PS will remain 3D. In contrast, using the SFB kernel in \cref{eq:Wnlq_kernel} and calculating the SFB PS with \cref{eq:dSFB_compute} effectively reduce the integral to 2D, which renders the computation more tractable.

We next show the inverse transform that maps the SFB back to the TSH space. Substituting \cref{eq:sfb_discrete_fourier_pair_a} into \cref{eq:TSH-x}, we have:
\begin{align}
\delta_{\ell m}(x)&=\int_{\hat{\bn}} Y_{\ell m}^*(\hat{\bn})\sum_{n\ell'm'}g_{n\ell'}(x)Y_{\ell'm'}(\hat{\bn})\delta_{n\ell'm'}\nonumber\\
&=\sum_{n}g_{n\ell}(x)\delta_{n\ell m}\,,
\label{eq:lmx-nlm}
\end{align}
where we applied the orthogonality of the spherical harmonics in \cref{eq:YY_ortho}. The angular power spectrum defined in \cref{eq:TSH} is then related to the SFB power spectrum defined in \cref{eq:SFB-discrete-PS} through:
\begin{align}
C_{\ell}(x,x')=\sum_{n_1,n_2}g_{n_1\ell}(x)g_{n_2\ell}(x')C_{\ell n_1n_2}\,,
\label{eq:TSH-SFB}
\end{align}
where the TSH sums over the Fourier modes present in the SFB. 

According to \cref{eq:TSH-SFB}, it is theoretically possible to calculate an arbitrary angular power spectrum through the SFB power spectrum when $x$ and $x'$ are within the radial range considered for the SFB PS. In reality, however, \cref{eq:TSH-SFB} is difficult to use for computing the density and Newtonian RSD terms, since they do not decrease for smaller Fourier scales (as seen in \cref{fig:DRSD_10_15}), which necessitates the inclusion of many $n_1$ and $n_2$ terms to account for higher $k_{n\ell}$ modes. The computational complexity of the SFB PS generally scales with $k_{\rm max}^3$, making it expensive to evaluate for smaller Fourier scales, so the mapping is difficult to use for the DRSD term. Fortunately, all GR terms substantially decrease when the Fourier mode $k_{n\ell}$ increases, as we discussed in \cref{sec:result}. As a result, computing GR effects in TSH through \cref{eq:TSH-SFB} only requires the inclusion of the SFB PS at lower Fourier modes, and we can obtain accurate results of GR effects in TSH at lower angular multipoles. This map from the SFB PS to TSH will be instrumental for validating our GR calculations in \cref{sec:validation}.

In addition to computing TSH through SFB, \cref{eq:TSH-SFB} also offers a clear indication where the TSH needs non-linear modeling, since the expression explicitly gives which Fourier $k_{nl}$ modes are folded into the TSH. For realistic galaxy surveys, it is impossible to estimate the TSH between arbitrary redshift pairs. One instead \HL{estimates} a projected version of TSH (pTSH) where the 3-D galaxy field is projected along the radial direction within some redshift bin. We give and discuss the mapping from SFB to pTSH in \cref{sec:pTSH-SFB}. 

To our knowledge, our work represents the first time a mapping from the discrete SFB power spectrum to TSH is explicitly written down and used for numerical calculations. An analog to \cref{eq:TSH-SFB} that maps the continuous SFB power spectrum to TSH can be similarly derived by substituting \cref{eq:sfb_x-to-k} into \cref{eq:TSH-x}:
\begin{align}
    C_{\ell}(x,x')=&\frac{4}{\pi^2}\int_{0}^{\infty}dk_1\,\int_{0}^{\infty}dk_2\,k_1^2k_2^2\nonumber\\
&\quad j_{\ell}(k_1x)j_{\ell}(k_2x')C_{\ell}(k_1,k_2)\,.
    \label{eq:TSH-cSFB}
\end{align}
However, it is significantly harder to apply the continuous version in \cref{eq:TSH-cSFB}, since one needs fine resolution of $k$ modes to perform integration over all positive $k$ values, while in the discrete case of \cref{eq:TSH-SFB}, the integration reduces to a sum over well-defined discrete $k_{nl}$ modes. This suggests another advantage of the discrete SFB basis: the transformations from the discrete SFB PS to other two-point statistics have relatively simple, stable numerical implementations.

\subsection{SFB-to-pTSH Mapping}\label{sec:pTSH-SFB}
We next consider the radially-integrated angular power spectrum, which we also refer to as the projected tomographic spherical harmonic (pTSH). Using the relationship between the spherical harmonic mode $\delta_{\ell m}(x)$ and the SFB mode $\delta_{n\ell m}$ of the unprojected density field in \cref{eq:lmx-nlm}, we have
\begin{align}
\delta_{\ell m}&=\int_{x_{\rm min}}^{x_{\rm max}}dx\, x^2 \delta_{\ell m}(x)\nonumber\\
&=\sum_{n}\left[\int_{x_{\rm min}}^{x_{\rm max}}dx\,x^2g_{n\ell}(x) \right]\delta_{n\ell m}\nonumber\\
&=\sum_n d_{n\ell} \delta_{n\ell m},
\end{align}
where 
\begin{align}
d_{n\ell}=\int_{x_{\rm min}}^{x_{\rm max}}dx\,x^2g_{n\ell}(x)
\,.
\end{align}
Therefore, the pTSH is related to the SFB PS as the following:
\begin{align}
C_{\ell}=\sum_{n,n'}d_{n\ell}d_{n'\ell} C_{\ell nn'}\,.
\label{eq:pTSH-SFB}
\end{align}
The above equation is the extension of \cref{eq:TSH-SFB} for the projected angular power spectrum measured in real surveys. This mapping can help to establish consistency between the SFB PS and pTSH measurements and explicitly show the exact radial Fourier modes being included at each angular multipole for a projected measurement.

\subsection{SFB-to-2CF Mapping}\label{sec:CF-SFB}
We now build upon \cref{eq:TSH-SFB} to obtain the mapping from the SFB PS to the two-point correlation function $\langle\delta(\bx)\delta(\bx')\rangle$. It is well-known that the two-point correlation function can be expanded with Legendre polynomials 
into the angular power spectrum \cite{13DiDio_classgal,21Zhang_SFB_TSH}:
\begin{equation}
\langle\delta(\bx)\delta(\bx')\rangle=\frac{1}{4\pi}\sum_{\ell}(2\ell+1)C_{\ell}(x,x')\mathcal{L}_\ell(\hat{\bn}\cdot\hat{\bn}')\,.
\label{eq:CF-TSH}
\end{equation}
Substituting \cref{eq:TSH-SFB} into \cref{eq:CF-TSH}, we have:
\begin{align}
    \langle\delta(\bx)\delta(\bx')\rangle=&\frac{1}{4\pi}\sum_{\ell,n_1,n_2}(2\ell+1)\mathcal{L}_\ell(\hat{\bn}\cdot\hat{\bn}')\nonumber\\
&\quad g_{n_1\ell}(x)g_{n_2\ell}(x')C_{\ell n_1n_2},
    \label{eq:2CF-SFB}
\end{align}
which can also be derived directly from \cref{eq:sfb_discrete_fourier_pair_a}. 

The above mapping can be useful for developing an optimal quadratic estimator for the SFB PS at the ultra-large scales. The existing estimator for the SFB PS developed in Ref.~\cite{21Gebhardt_SuperFab} is based on the pseudo-$C_{\ell}$ approach~\cite{01Wandelt_Cl,19Alonso_Nmaster}, which performs well for small angular scales but becomes non-optimal for the largest angular scales. The optimal quadratic estimator requires knowing the covariance between any two galaxy positions (or pixels) and the derivative of the covariance with respect to the band-powers~\HL{\cite{98Tegmark_quadratic_PS,21Philcox_quadratic_PS,08Hamilton_PS}}. \cref{eq:2CF-SFB} directly expresses the covariance between any two galaxy positions (the correlation function) in terms of the band-powers (the SFB PS), so it can be used for building a quadratic estimator for the SFB PS\footnote{\HL{See Refs.~\cite{00Hamilton_PKL_PSCz,02Tegmark_PKL_2df,08Hamilton_PS} for using the “pseudo-Karhunen-Lo\`eve” (PKL) modes, the product of logarithmic spherical waves $e^{i\omega \ln x}$ and spherical Harmonics $Y_{\ell m}(\hat{\bx})$, in the quadratic estimator for galaxy clustering. These PKL modes are similar to the SFB modes except using a different set of radial basis functions.}}.

The correlation function can be reparametrized in terms of the relative displacement $\bs=\bx-\bx'$ between the two galaxies:
\begin{equation}
\xi(s,x,\hat{\bn}\cdot\hat{\bs})=\xi(\bs,\bx)\equiv\langle\delta(\bx)\delta(\bx-\bs)\rangle\,.
\end{equation}
It can then be decomposed with respect to $\mu\equiv \hat{\bn}\cdot\hat{\bs}$ to obtain the the two-point correlation function multipoles (2CFM) under the end-point LOS:
\begin{align}
    \xi_{L}(s,x)\equiv\frac{2L+1}{2}\int_{\mu=-1}^{1}\mathcal{L}_{L}(\mu) \xi(s,x,\mu)\,,
\end{align}
which are frequently used in the analysis of spectroscopic surveys. 

One can then express the 2CFM as a map from the SFB PS:
\begin{align}
    \xi_{L}(s,x)\equiv\frac{2L+1}{8\pi}\sum_{\ell,n_1,n_2}(2\ell+1)g_{n_1\ell}(x)C_{\ell n_1n_2}S_{n_2\ell}^{L}\,,
    \label{eq:2CFM-SFB}
\end{align}
where
\begin{align}
    S_{n_2\ell}^{L}\equiv&\int_{\mu=-1}^{1}\mathcal{L}_{L}(\mu) \mathcal{L}_\ell(\hat{\bn}\cdot\hat{\bn}')\nonumber\\
&\quad g_{n_2\ell}(x'=\sqrt{x^2+s^2-2xs\mu})\,.
\end{align}

Therefore, another possible validation route for the SFB calculation performed in this work is to compare the 2CFM results transformed from our SFB calculation with existing codes \texttt{COFFE} \cite{18Tansella_coffe} or \texttt{GaPSE} \cite{23Foglieni} that can calculate the GR effects in the configuration-space 2CFM. However, the mapping from the SFB PS to the 2CFM in \cref{eq:2CFM-SFB} is more complicated than the SFB-to-TSH map in \cref{eq:TSH-SFB}. In addition, \texttt{CLASS} is a more established and validated cosmological code, with two independent methods implemented for calculating the GR effects in TSH~\cite{13DiDio_classgal,18Schoneberg_cl_fftlog_gr}. Therefore, we choose the TSH over the 2CFM for validation in this work.

\section{Spherical Harmonic Decomposition} \label{sec:TSH-decomp}
\subsection{SFB Kernels}\label{sec:SFB-kernel}
In this appendix, we will derive \cref{eq:dSFB_compute,eq:Wnlq_kernel} that are used to compute the SFB power spectrum. Our derivation will rely on the TSH-to-SFB mapping given in \cref{eq:dSFB-TSH}.

Suppose we can write the spherical harmonic mode of the observed galaxy density in the following form:
\begin{align}
    \delta_{\ell m}(x)&=\frac{1}{2\pi^2}i^{\ell}\int_{\bq}\Delta_{\ell}(x,q)D_{\rm m,0}(\bq)Y_{\ell m}^{*}(\hat{\bq})\label{eq:angular-transfer},
\end{align}
where $\Delta_{\ell}(x,q)$ is the angular kernel and $D_{{\rm m},0}$ is the comoving matter density at the present time. We will show in the following section that all GR effects can indeed be expressed into the above form in terms of some angular kernels. The angular power spectrum (TSH) defined in \cref{eq:TSH} becomes:
\begin{align}
&\langle\delta_{\ell m}(x)\delta_{\ell' m'}^{*}(x')\rangle=\frac{1}{4\pi^4}i^{\ell-\ell'}\int_{\bq}\Delta_{\ell}(x,q)Y_{\ell m}^{*}(\hat{\bq})\nonumber\\
&\qquad \int_{\bq'}\Delta_{\ell'}(x',q')Y_{\ell' m'}(\hat{\bq}')\langle D_{\rm m,0}(\bq)D_{\rm m,0}(\bq')\rangle\,.\label{eq:TSH-deriv-int}
\end{align}

The translational invariance and the statistical isotropy of the universe for the matter distribution imply that
\begin{align}
    \langle D_{\rm m,0}(\bq)D_{\rm m,0}(\bq')\rangle=(2\pi)^3\delta(\bq-\bq')P_{\rm m,0}(q),
\end{align}
where $P_{\rm m,0}(q)$ is the matter power spectrum at the present time. Substituting the above matter power spectra into \cref{eq:TSH-deriv-int}, we have
\begin{align}
   &C_{\ell}(x(z),x'(z'))=\frac{2}{\pi}\int_{0}^{\infty}dq\,q^2\Delta_{\ell}(x,q)\Delta_{\ell'}(x',q)\nonumber\\
&\qquad P_{\rm m,0}(q)i^{\ell-\ell'}\int_{\hat{\bq}}Y_{\ell m}^{*}(\hat{\bq})Y_{\ell' m'}(\hat{\bq})\,.
\end{align}
We then use the orthogonality of the spherical Harmonics in \cref{eq:YY_ortho} to obtain:
\begin{equation}
C_{\ell}(x,x')=\frac{2}{\pi}\int_{0}^{\infty}dq\, q^2 P_{\rm m,0}(q)\Delta_{\ell}(x,q)\Delta_{\ell}(x',q)\,,
\label{eq:Cl-expression}
\end{equation}
which is the well-known formula that relates the galaxy angular power spectrum to the underlying matter power spectrum at the linear order and is used for the numerical computation of the TSH. 

Substituting the above expression \cref{eq:Cl-expression} into the TSH-to-SFB map in \cref{eq:dSFBR-TSH}, we have
\begin{align}
    C_{\ell n_1n_2}^{\rm R}&=\frac{2}{\pi}\int_{x_{\rm min}}^{x_{\rm max}}dx\,x^2g_{n_1\ell}(x)R(x)\int_{x_{\rm min}}^{x_{\rm max}}dx'\,x'^2 g_{n_2\ell}(x')\nonumber\\
&\qquad R(x')\int_{0}^{\infty}dq\, q^2 P_{\rm m,0}(q)\Delta_{\ell}(x,q)\Delta_{\ell}(x',q)\,.\nonumber
\end{align}
We can then rearrange the integration order between $x,x'$ and $q$ to obtain the formulas for the SFB PS and SFB kernels shown in \cref{eq:dSFB_compute,eq:Wnlq_kernel}:
\begin{align}
    C_{\ell n_1n_2}^{\rm R}&=\int_{0}^{\infty}dq\, \mathcal{W}_{n_1\ell}^{\rm R}(q) \mathcal{W}_{n_2\ell}^{\rm R}(q) P_{\rm m,0}(q)\label{eq:dSFB_compute-appendix}\,,\\
    \mathcal{W}_{n\ell}^{\rm R}(q)&\equiv\sqrt{\frac{2}{\pi}}q\int_{x_{\rm min}}^{x_{\rm max}}dx\,x^2g_{n\ell}(x)R(x)\Delta_{\ell}(x,q)\label{eq:Wnlq_kernel-appendix}\,.
\end{align}

Following the same process, we can write down the analog of \cref{eq:dSFB_compute-appendix} for the continuous SFB power spectrum defined in \cref{eq:SFB-PS}:
\begin{align}
    C_{\ell}^{\rm R}(k_1,k_2)&=\int_{0}^{\infty}dq\, \mathcal{W}_{\ell}^{\rm R}(k_1,q) \mathcal{W}_{\ell}^{\rm R}(k_2,q) P_{\rm m,0}(q)\label{eq:cSFB_compute}\,,
\end{align}
where the continuous SFB kernel is
\begin{align}
\mathcal{W}_{\ell}^{\rm R}(k,q)\equiv\sqrt{\frac{2}{\pi}}q\int_{0}^{\infty}dx\,x^2j_{\ell}(kx)R(x)\Delta_{\ell}(x,q)\label{eq:Wkq_kernel}\,.
\end{align}
The continuous SFB transform is performed over the entire configuration space, which is impossible for numerical implementation. It is only possible to compute the CSFB PS under some radial selection function $R(x)$ such that the integration range in \cref{eq:Wkq_kernel} becomes finite. We note that the continuous and discrete SFB power spectra share the same numerical structure except using different radial basis functions in the SFB kernels, so our numerical approach described in \cref{sec:compute-numerics} will work for both cases.

\subsection{GR Angular Kernels} \label{sec:GR_kernel}
We next show all GR terms for the galaxy number count in \cref{eq:GR} can indeed be put into the form of \cref{eq:angular-transfer}, that is we will review the derivation of the GR angular kernels summarized in \cref{sec:compute-analytic}. 

For any scalar quantity $A(\bx)$, using the Rayleigh's wave expansion \cref{eq:planewave-Ylm} and the corresponding linear transfer function $T_{A}(x,q)$ for the scalar, we have
\begin{align}
    A(\hat{\bn},x(z))&=\frac{1}{(2\pi)^3}\int_{\bq} A(x,\bq)e^{i\bq\cdot\bx}\nonumber\\
    &=\frac{1}{2\pi^2}\sum_{\ell m}i^{\ell}\int_{\bq}T_{A}(x,q)D_{m,0}(\bq)\nonumber\\
    &\qquad j_{\ell}(qx)Y_{\ell m}^{*}(\hat{\bq})Y_{\ell m}(\hat{\bn})\,.
\end{align}
From the above expression, we can read off the spherical harmonic decomposition for the scalar:
\begin{align}
    A_{\ell m}(x)&=\frac{i^{\ell}}{2\pi^2}\int_{\bq}T_{A}(x,q)D_{m,0}(\bq)j_{\ell}(qx)Y_{\ell m}^{*}(\hat{\bq})\,,
\end{align}
and we can identify the angular kernel as
\begin{align}
    \Delta_{\ell}^{A}(x,q)=T_{A}(x,q)j_{\ell}(qx)\,.\label{eq:D_scalar}
\end{align}
The above \cref{eq:D_scalar} applies to the density and the NIP terms, resulting in the first half of \cref{eq:D_R} and \cref{eq:D_G}.

For an integrated scalar quantity, we have
\begin{align}
    \int_{0}^{x}dr\, A(\hat{\bn},r)&=\frac{1}{(2\pi)^3}\int_{\bq}  \int_{0}^{x}dr\,A(r,\bq)e^{i\bq\cdot r\hat{\bn}}\nonumber\\
    &=\frac{1}{2\pi^2}\sum_{\ell m}i^{\ell}\int_{\bq}\int_{0}^{x}dr\,T_{A}(r,q)D_{m,0}(\bq)\nonumber\\
    &\qquad j_{\ell}(qr)Y_{\ell m}^{*}(\hat{\bq})Y_{\ell m}(\hat{\bn})\,.
\end{align}
Therefore, the angular kernel for an integrated term is 
\begin{align}
    \Delta_{\ell}^{\int A}(x,q)=\int_{0}^{x}dr\,T_{A}(r,q)j_{\ell}(qr)\label{eq:D_int_scalar}\,,
\end{align}
which can be used for the Shapiro and ISW effects to obtain \cref{eq:D_S,eq:D_I} and also for the lensing potential $\psi^{\rm lens}$ (defined in \cref{eq:kappa}). For the lensing convergence $\kappa =\frac{1}{2}\nabla^2_{\hat{\bn}'}\psi^{\rm lens}$, the angular Laplacian becomes multiplication in the harmonic space:
\begin{align}
    \kappa_{\ell m}=-\frac{1}{2}\ell(\ell+1)\psi^{\rm lens}_{\ell m},
\end{align}
which will lead to the angular kernel \cref{eq:D_L} for the lensing convergence. 

For the RSD and Doppler terms that contain the LOS velocity, we note the following relationship between the LOS velocity $\hat{\bn}\cdot\bv$ and the velocity scalar $v$ in the Fourier space:
\begin{align}  
\hat{\bn}\cdot\bv(k) e^{i\bk\cdot\hat{\bn}x}&=-i\hat{\bn}\cdot\bk V(k)e^{i\bk\cdot\hat{\bn}x}\nonumber\\
&=-v(k)\partial_{kx}e^{i\bk\cdot\hat{\bn}x}\,.
\end{align}
Therefore,
\begin{align}
\hat{\bn}\cdot\bv(\hat{\bn},x)&=-\frac{1}{2\pi^2}\sum_{\ell m}i^{\ell}\int_{\bq}T_{v}(x,q)D_{m,0}(\bq)\nonumber\\
&\qquad j_{\ell}'(qx)Y_{\ell m}^{*}(\hat{\bq})Y_{\ell m}(\hat{\bn})\,,
\end{align}
and we obtain the angular kernel for the LOS velocity:
\begin{align}
\Delta_{\ell}^{\vec{n}\cdot\vec{v}}(x, q)=-T_{v}(x,q)j_{\ell}'(qx)\label{eq:D_v},
\end{align}
which will lead to the angular kernels for the Doppler term in \cref{eq:D_D} and the RSD term in the second half of \cref{eq:D_R}.

\subsubsection{Observer's Terms}\label{sec:observer}

We now give the angular kernels for the observers' terms in \cref{eq:GR} at the linear order. For the observer's potential, we can simply apply \cref{eq:D_scalar} at $x=0$. Since $j_{\ell}(0)=\delta_{\ell0}$, the observer's potential only impacts the monopole, and the angular kernel becomes:
\begin{align}
\Delta_{0}^{P_o}(x,q)&=\Bigg[\left(\mathcal{A}_1(x)\mathcal{H}_0-\frac{2-5s(x)}{x}\right)V(0,q)\nonumber\\
&\qquad -\mathcal{A}_1(x)\Psi(0,q)\Bigg]\,.\label{eq:D_obs_p}
\end{align}
We can apply \cref{eq:D_v} to the observer's velocity at $x=0$. Since $j'_\ell(0)=\frac{1}{3}\delta_{\ell1}$, the observer's velocity only impacts the dipole and has the following angular kernel:
\begin{align}
\Delta_{1}^{v_o}(x,q)&=-\frac{1}{3}\left(\mathcal{A}_1(x)+2-5s(x)\right)v(0,q)\,.\label{eq:D_obs_v}
\end{align}
This completes the derivation of the angular kernels for all linear-order terms present in the relativistic galaxy number count.

\begin{figure*}[tbp]
\centerline{\includegraphics[width=0.8\textwidth]{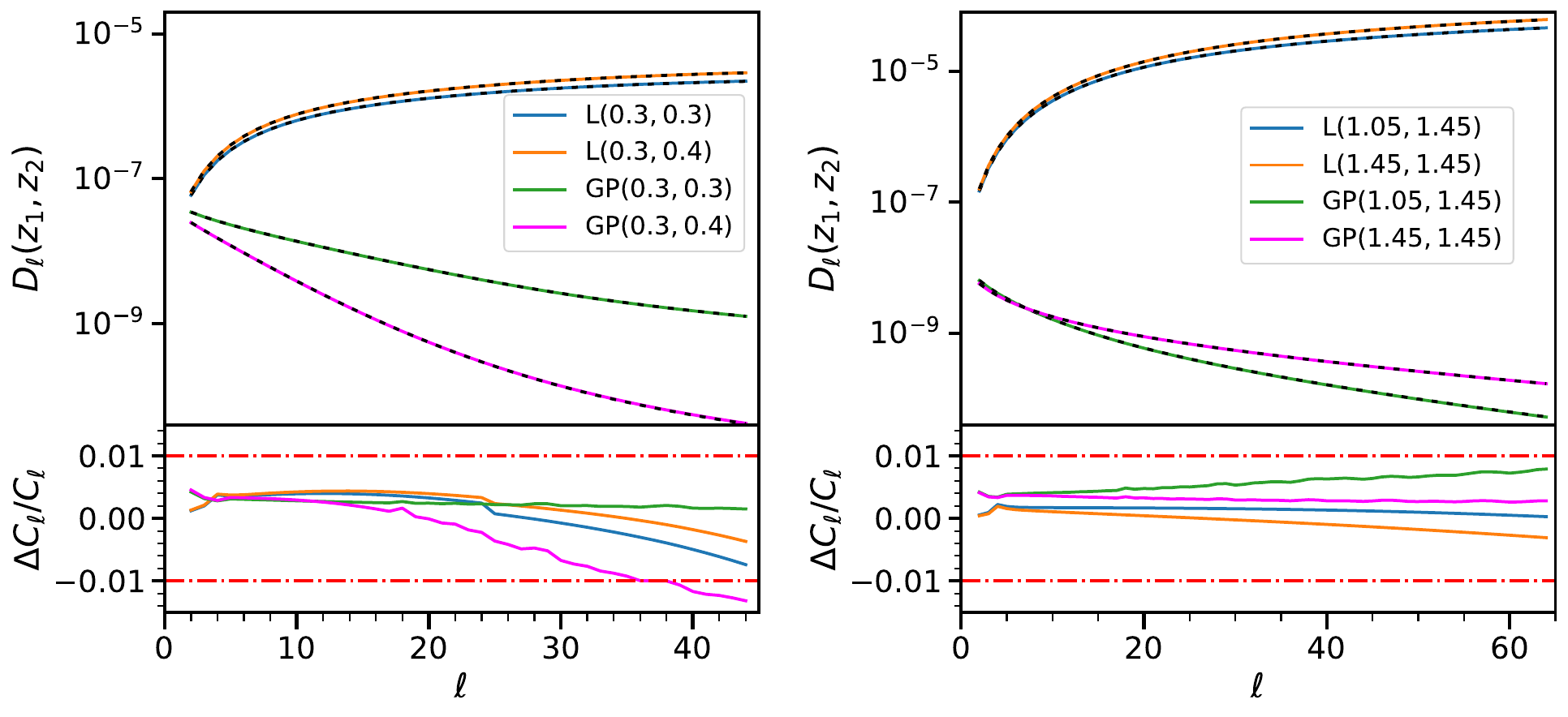}}
\caption{The multipole-scaled angular power spectrum $D_{\ell}\equiv\ell(\ell+1)/(2\pi)C_{\ell}$ and the relative difference $(C_{\ell}^{\rm SFB}-C_{\ell}^{\rm class})/C_{\ell}^{\rm class}$ between our SFB-transformed results and the \texttt{CLASS} outputs. In the upper panels, the solid lines represent the \texttt{CLASS} output, while the dashed black lines show the corresponding angular power spectrum transformed from our SFB PS results. In both panels, the different colors indicate the different GR effects and different redshifts considered. In the legend, \HL{``L"} indicates the lensing-lensing term, while \HL{``GP"} indicates the GP-GP term that includes Shapiro, ISW, and NIP effects. The two numerical values inside the bracket indicate the two redshifts where the angular power spectrum is computed. The left and right subplots consider the lower ($z=0.2-0.5$) and the higher ($z=1.0-1.5$) redshift ranges respectively.}
\label{fig:valid}
\end{figure*}

\section{Validation}\label{sec:validation}

The computation of the SFB PS for the Newtonian terms (including the standard density plus RSD and Newtonian Doppler terms given in \cref{eq:Newtonian}) have already been validated with lognormal mocks in Refs.~\cite{23Gebhard_SFB_eBOSS} and \cite{24PSM_SFB}. The goal here is to validate the SFB implementation for the other GR terms, especially the integrated terms such as lensing of which the numerical computation is non-trivial. 

A mock-based validation of GR effects will require implementing the linear-order relativistic effects in \cref{eq:GR} on a large number of N-body or lognormal simulations with large volumes \cite{17Borzyszkowski_LIGER,22Breton_Magrathea,20Leopori_lensing_ray_tracing}, which is a rather complicated process. 
Therefore, it is better and simpler to validate our implementation of GR effects with theoretical computation. Our validated theoretical computation can then be used to validate GR simulations and perform inference in the upcoming LSS surveys. 

Our validation of GR effects in SFB will rely on the SFB-to-TSH mapping in \cref{eq:TSH-SFB}. We can transform our SFB results into TSH using \cref{eq:TSH-SFB} and then compare with the TSH results directly calculated from \texttt{CLASS}. The number count angular power spectrum in \texttt{CLASS} has been implemented with both brute-force integration \cite{13DiDio_classgal} and FFTlog method \cite{18Schoneberg_cl_fftlog_gr}, with both methods agreeing at $0.1\%$ for the angular PS of the total relativistic number count. Our numerical implementation of the SFB power spectrum is completely independent \HL{of} \texttt{CLASS}, and the angular power spectrum is only obtained as the last step from our SFB calculation using the mapping such that we can compare with the existing public code on GR effects. 

In the summation of \cref{eq:TSH-SFB}, we consider $n_1$ and $n_2$ with $k_{n\ell}\leq 0.25 h/{\rm Mpc}$ to truncate the smaller radial modes. We consider the TSH obtained from both auto-correlation at one redshift and cross-correlation between two redshifts. The bias parameters are set as the same ones chosen in \cref{sec:result}. We show our validation results in \cref{fig:valid} for the lensing-lensing and GP-GP terms\footnote{Note that the velocity potential $\left(b_{\rm e}(x)-3\right)\mathcal{H}(x)V(x, q)$ present in our NIP term (the fourth line of \cref{eq:GR}, corresponding to the angular kernel in \cref{eq:D_G}) is classified as part of the Doppler term in the \texttt{CLASS} convention (see Appendix A of Ref.~\cite{13DiDio_classgal}). In order to match the \texttt{CLASS} results, we remove the velocity potential from the GP term during validation.} at different redshifts. 

In general, our SFB-transformed TSH results for the two individual GR terms agree within the percent level of the corresponding \texttt{CLASS} outputs at both low and high redshifts (the left and right subplots of \cref{fig:valid}) for multipoles $2\leq\ell\lessapprox 40$. This demonstrates the accuracy of our implementation of the lensing and GP terms in the SFB PS. Since the calculation of the Newtonian RSD and Doppler terms have already been validated through mocks in Refs.~\cite{23Gebhard_SFB_eBOSS,24PSM_SFB}, and the GP term contains the NIP, Shapiro, and ISW effects, we have therefore validated the SFB calculation for all terms present in the relativistic number count.

Our SFB-transformed TSH begins to diverge from the \texttt{CLASS} outputs, gradually surpassing the percentage levels for higher multipoles $\ell \gtrapprox 60$ as seen in \cref{fig:valid}. This deviation is expected, since one has to include more radial SFB modes evaluated at higher $k_{n\ell}$ than our truncation limit $0.25 h/{\rm Mpc}$ to obtain more accurate TSH for higher angular multipoles. Since \texttt{CLASS} uses the full linear transfer function from solving the Boltzmann equations, while our SFB computation relies on the $\Lambda$CDM approximations given in \cref{eq:approx-D,eq:approx-v,eq:approx-Phi,eq:approx-dot-Phi}, in addition to the errors introduced by the $k_{n\ell}$ truncation in the TSH-SFB mapping, we expect pushing our validation below the percent-level agreement to be difficult. However, the total GR effect only produces percent-level correction (below $10\%$ as seen in \cref{fig:GR_term_evolve}) at redshift $z\leq 2$, we consider achieving percent-level agreement on the GR corrections sufficient for validating the accuracy of our implementation.

\section{Lensing Approximation}\label{sec:lens-approx}

\subsection{Limber Approximation on LOS Integral}
In this section, we aim to gain some intuition for the lensing term in the SFB power spectrum by employing the Limber approximation on the LOS integral in \cref{eq:kappa}. At the first-order, the Limber approximation is \cite{08Loverde_limber}:
\begin{align}
\label{eq:jl_limber}
j_\ell(kr)
&\simeq \sqrt{\frac{\pi}{2rk}}\frac{1}{k}\delta^D(r-\frac{\ell+\frac{1}{2}}{k})\,.
\end{align}
The above approximation applies when all other
functions are slowly-varying compared to the frequency
of the spherical Bessel functions, and the integration should be over
a wide interval.

Applying the Limber approximation to the lensing angular kernel in \cref{eq:D_L}, we have
\begin{align}
&\Delta_{\ell}(x, q)^{\rm Lensing}\approx\ell(\ell+1)\frac{2-5 s(x)}{2}\int_{r=0}^{\infty}\mathbb{I}_{[r\leq x]}\frac{x-r}{x r}\nonumber\\
&\qquad(\Phi(r, q)+\Psi(r, q))\sqrt{\frac{\pi}{2rq}}\frac{1}{q}\delta^D(r-\frac{\ell+\frac{1}{2}}{q})\nonumber\\
&=\sqrt{\frac{\pi}{2(\ell+\frac{1}{2})}}\ell(\ell+1)\frac{2-5 s(x)}{2}\left(\frac{1}{\ell+\frac{1}{2}}-\frac{1}{qx}\right)\nonumber\\
&\qquad\left(\Phi(\frac{\ell+\frac{1}{2}}{q}, q)+\Psi(\frac{\ell+\frac{1}{2}}{q}, q)\right)\mathbb{I}_{[(\ell+\frac{1}{2})/q\leq x]}\label{eq:limber_lensing}\,,
\end{align}
where $\mathbb{I}_{[...]}$ is the indicator function. We see that the LOS integral is mostly determined by the gravitational potentials at an effective position $x_{\rm lens}=(\ell+\frac{1}{2})/q$ for each combination of the Fourier mode $q$ and angular mode $\ell$. The source galaxy at $x$ is effectively lensed by the structure at $x_{\rm lens}$ for a given SFB mode. If the Fourier scale $q$ is fixed, then increasing the angular scale will shift the effective lens plane to a higher redshift.

\begin{figure*}[tbp]
\centerline{\includegraphics[width=0.75\textwidth]{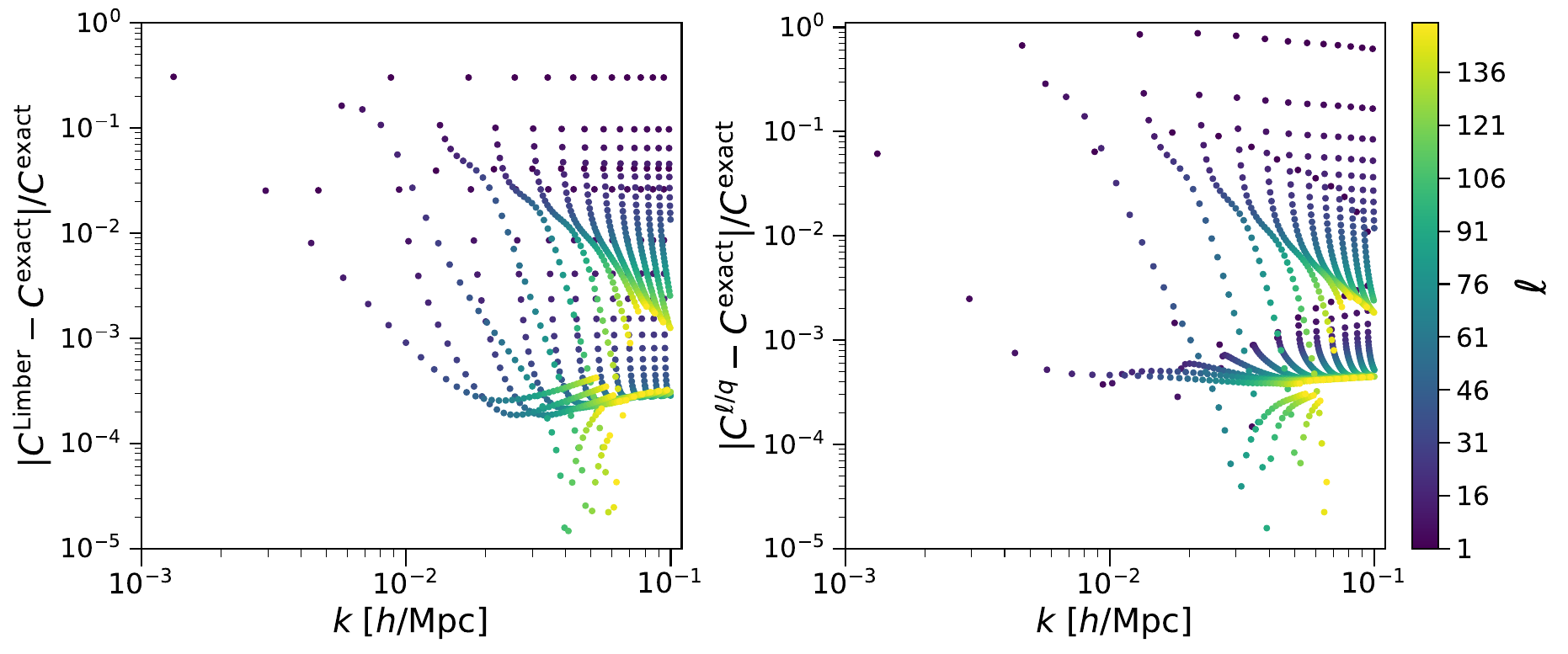}}
\caption{The accuracy of two approximations for the lensing-lensing term in the diagonal part of the SFB power spectrum for the redshift range from $z=1.0$ to $1.5$. The left panel shows the accuracy of \cref{eq:limber_lensing}, where the Limber approximation is applied to the LOS integral in the lensing angular kernel. The right panel shows the accuracy of \cref{eq:l-q-approximation}, where the explicit dependence of the lensing angular kernel on $\ell$ and $q$ are eliminated. The accuracy of both methods primarily \HL{depends} on the angular multipoles $\ell$, and they agree with the exact calculation at the percent level for $\ell\gtrapprox 50$.}
\label{fig:approx_lens_10_15}
\end{figure*}

As shown in the left panel of \cref{fig:approx_lens_10_15}, the limber approximation deviates from the exact results of the lensing-lensing term by tens of percent at the largest angular scales across the Fourier modes $k$, but it becomes accurate within the percent level for higher angular multipoles. This potentially suggests that one can use the Limber approximation to further speed up the evaluation of the lensing effect in the SFB power spectrum at smaller angular scales by avoiding the LOS integral. However, the exact applicability of the approximation will depend on the accuracy required by the specific science goal, and it should be examined for each survey. We note that the Limber approximation on the LOS integral also applies to other integrated effects including the Shapiro and ISW effects.

\subsection{$\ell/q$-Approximation}
Here we will examine the accuracy of the following approximation for the lensing angular kernel:
\begin{align}
\Delta_{\ell}(x, q)^{\rm Lensing}&\approx \frac{3\Omega_{\rm m,0}H_0^2}{2}\left(2-5 s(x)\right)\int_{0}^{x}dr\,r\frac{r-x}{x}\nonumber\\
&\qquad(1+z(r))D(r)j_{\ell}(qr)\label{eq:l-q-approximation}\,.
\end{align}
The above approximation for the lensing magnification is often used for calculating the angular power spectrum in literature~\cite{08Loverde_lensing_TSH,16Dizgah_Eg,18Yang_EG,24Krolewski_CMB_quadar_fnl}, and it has been used for computing the SFB PS in Ref.~\cite{21Zhang_SFB_TSH}. 

To see how \cref{eq:l-q-approximation} can be obtained, we substitute the density and gravitational potential transfer functions \cref{eq:approx-D,eq:approx-Phi} under the $\Lambda$CDM model into the exact expression of \cref{eq:D_L}:
\begin{align}
    &\Delta_{\ell}(x, q)^{\rm Lensing}=\frac{3}{2}\frac{\ell(\ell+1)}{q^2}\left(2-5 s(x)\right)\int_{0}^{x}dr\,\frac{r-x}{x r}\nonumber\\
&\qquad\mH(r)^2\Omega_{\rm m}(r)D(r)j_{\ell}(qr)\nonumber\\
    &=\frac{3\Omega_{\rm m,0}H_0^2}{2}\left(2-5 s(x)\right)\int_{0}^{x}dr\,\frac{\ell(\ell+1)}{q^2}\frac{r-x}{xr}\nonumber\\
&\qquad(1+z(r))D(r)j_{\ell}(qr)\label{eq:D_L_CDM}\,.
\end{align}
Motivated by the Limber approximation, we can approximate the $\ell(\ell+1)/q^2$ factor in \cref{eq:D_L_CDM} as $r^2$ to obtain \cref{eq:l-q-approximation}. We refer to this scheme as the \HL{``$\ell/q$-approximation"}, since the explicit dependence of the lensing angular kernel on $\ell$ and $q$ has been dropped. 

Similar to the Limber approximation applied to the LOS integral, the $\ell/q$-approximation is inaccurate at the largest angular scales and becomes more accurate for the lensing SFB PS when one considers higher $\ell$ values, as shown in the right panel of \cref{fig:approx_lens_10_15}. Despite the accuracy of this approximation, it does not achieve significant computational speedup for evaluating the SFB power spectrum, since the LOS integral remains in the angular kernel, and the dimension of the integral under the approximation is the same as the exact expression. We therefore recommend using the exact angular kernel \cref{eq:D_L} when computing the SFB PS.

\section{Infrared Divergence in Monopole}\label{sec:divergence}

We have so far computed the SFB power spectrum for angular multipoles $\ell\geq 1$. We now discuss the computation of the SFB monopole and observe its divergence for angular kernels of the form:
\begin{align}
    \Delta_{\ell}^{\rm Div}(x,q)=\frac{1}{q^2}j_{\ell}(qx)A(x)\,,
    \label{eq:div-kernel}
\end{align}
where $A(x)$ absorbs both the radial window and redshift evolution. For the DRSD and Doppler terms, their monopole is convergent. The lensing term does not contribute to the monopole since its angular kernel is proportional to $\ell(\ell+1)$ as seen in \cref{eq:D_L}. In comparison, the angular kernels for the NIP, Shapiro, ISW, and PNG terms have the same $1/q^2$ behavior as in \cref{eq:div-kernel} due to their proportionality to the gravitational potential, and the correlation among these terms will diverge for the SFB monopole in both the continuous and discrete SFB basis.

\subsection{Divergence in Discrete SFB}
For the discrete SFB power spectrum computed in \cref{eq:dSFB_compute}, we have
\begin{align}
&C_{\ell nn'}^{\rm Div}=\frac{2}{\pi}\int_{0}^{\infty}dq\,\frac{1}{q^2}\left[\int_{x_{\rm min}}^{x_{\rm max}}dx\, x^2 g_{n\ell}(x)j_{\ell}(qx)A(x)\right]\nonumber\\
&\qquad\left[\int_{x_{\rm min}}^{x_{\rm max}}dx'\, x'^2 g_{n'\ell}(x')j_{\ell}(qx')A(x')\right]P_{\rm m,0}(q)\,.
\label{eq:div-SFB_monopole_discrete}
\end{align}
In general, the spherical Bessel function has the asymptotic behavior $j_{\ell}(y)\sim O(y^{\ell})$ for $y\to0$. We see that for higher multipoles $\ell\geq 1$, the integrand behaves as $q^{2\ell-2}P_{\rm m,0}(q)$, which will lead to convergence in the above integral. Therefore, there is no divergence for any GR terms in SFB multipoles $\ell\geq 1$.

For the monopole $\ell=0$, the integrand now scales as $q^{-2}P_{\rm m,0}(q)\sim q^{-2+n_s}$, where $n_s$ is the scalar index for the primordial perturbation spectrum. Since the measurement of CMB anisotropies has strongly constrained the scalar index to be $n_s<1$ \cite{18Planck_Parameter}, the scaling behavior of $q^{-2+n_s}$ will cause divergence in \cref{eq:div-SFB_monopole_discrete}, that is $C_{0nn'}$ diverge for all $n$ and $n'$ regardless of the boundary conditions. Therefore, the correlation among the NIP, Shapiro, ISW, and PNG terms will be divergent for the discrete SFB monopole at all Fourier modes. 

\subsection{Divergence in Continuous SFB}
When $A(x)$ is regulated by some radial selection function with finite coverage from $x_{\rm min}$ to $x_{\rm max}$, the continuous SFB monopole from \cref{eq:cSFB_compute} becomes:
\begin{align}
&C_{0}^{\rm R,Div}(k,k')=\frac{2}{\pi}\int_{0}^{\infty}\frac{dq}{q^2}\left[\int_{x_{\rm min}}^{x_{\rm max}}dx\,x^2 j_0(kx)j_{0}(qx)A(x)\right]\nonumber\\
&\quad\left[\int_{x_{\rm min}}^{x_{\rm max}}dx'\, x'^2 j_0(k'x')j_{0}(qx')A(x')\right]P_{\rm m,0}(q)
\label{eq:cSFB-div}
\end{align}
Similar to the above arguments in the discretized case, all $C_{0}(k,k')$ diverges regardless of the Fourier modes, since the integral $\int_{x} x^2 j_0(kx)j_{0}(qx)A(x)$ will always be finite when we integrate over some finite range, which does not help to tame the divergence caused by $q^{-2}$.

We now consider an idealized case of the infinite universe. Assuming that $A(x)=A$ is uniform across the entire configuration space without the presence of any window, we see that for all angular multipoles including the monopole, the CSFB PS of the divergent terms becomes
\begin{align}
&C_{\ell}^{\rm Div}(k,k') =\frac{2A^2}{\pi}\int_{0}^{\infty}dq\,\frac{1}{q^2}\left[\int_{0}^{\infty}dx\, x^2 j_{\ell}(kx)j_{\ell}(qx)\right]\nonumber\\
&\qquad\left[\int_{0}^{\infty}dx'\, x'^2 j_{\ell}(k'x')j_{\ell}(qx')\right]P_{\rm m,0}(q)\nonumber\\
&=\frac{2A^2}{\pi}\int_{0}^{\infty}dq\,\frac{1}{q^2}\frac{\pi}{2kq}\delta^{\rm D}(k-q)\frac{\pi}{2k'q}\delta^{\rm D}(k'-q)P_{\rm m,0}(q)\nonumber\\
&=\frac{\pi A^2}{4}\frac{1}{k^6}P_{\rm m,0}(k)\delta^{\rm D}(k-k'),
\label{eq:ideal-cSFB-div}
\end{align}
using the orthogonality of spherical Bessel functions in  \cref{eq:O_jl}. Therefore, in the case where one ignores the radial window function, the $k^{-4}$ scaling behavior for the divergent terms is exactly recovered in the CSFB PS\footnote{An additional $k^{-2}$ scaling factor present in \cref{eq:ideal-cSFB-div} is due to our choice of splitting the $2k^2/\pi$ factor in the continuous SFB decomposition given by \cref{eq:sfb_x-to-k,eq:sfb_k-to-x}. In an isotropic and homogeneous universe without any Newtonian RSD or GR effects, we expect $C_{\ell}(k,k')\sim k^{-2}P_{\rm m}(k)$.}, and we can see that the CSFB monopole converges at all Fourier modes except $k=0$. 

The behavior of \cref{eq:ideal-cSFB-div} in the infinite universe also explains the divergence of the CSFB monopole at all Fourier modes in the presence of radial window function in \cref{eq:cSFB-div}. In the infinite case, only $C_0(0,0)$ diverges, but the $k=0$ mode in the infinite case is mixed to all other $k$ modes after the window convolution, which causes the divergence of the CSFB monopole $C_0^{\rm R}(k,k')$ with radial window at all Fourier modes.

\subsection{Practical and Theoretical Considerations}
Fortunately, the divergence of the SFB monopole in individual GR terms is practically irrelevant for galaxy surveys due to the presence of integral constraint~\cite{19Mattia_IC} (also known as local average effects ~\cite{23Gebhard_SFB_eBOSS,20Wadekar_covariance}), i.e. the true average galaxy density is unknown but estimated from the survey itself. Without the presence of any angular mask (that is a window function with only radial selection), the measured SFB monopole at all Fourier modes is forced to be zero due to the radial integral constraint~\cite{23Gebhard_SFB_eBOSS,24gSFB}, and the cosmological SFB monopole becomes un-observable. Therefore, the divergence of the SFB monopole for the GP and PNG terms is of no concern when analyzing galaxy surveys.

However, the divergence of the SFB monopole is still of theoretical interest. Even though the individual NIP, ISW, and Shapiro effects are divergent in the SFB monopole, the divergence in the GP term will cancel out once the observer's potential, which we have ignored in this work, is consistently included. Ref.~\cite{23Mitsou_IR_Div} has argued for the general cancellation of the infrared-divergent terms under a gauge-invariant linear-order GR calculation with adiabatic conditions, while Refs.~\cite{20YooGR-P}, \cite{21Ginat_monopole}, and \cite{22CatorinaGR-P} have explicitly shown the cancellation of the divergent linear-order GR terms for the Fourier-space galaxy power spectrum, the angular monopole, and the configuration-space correlation function respectively. We will explicitly show the divergence cancellation and calculate the SFB monopole under the linear-order GR effects including the observer's terms in a forthcoming work.

In contrast to the GR terms where infrared divergence will cancel out, the PNG-PNG term (proportional to $f_{\rm NL}^2$) is genuinely divergent in the SFB monopole. Though initially formulated in the Newtonian theory, the scale-dependent bias due to the local PNG is consistent with the relativistic context as we have reviewed in \cref{sec:PNG}. Such divergence is in fact a signature of the local PNG effect, since it is sensitive to the super-sample value of the gravitational potential due to the coupling of the large and small scales during the structure formation. Despite the apparent degeneracy between the GP and PNG terms in the $\ell$-$k$ space as shown in \cref{fig:PNG_GP_compare}, the divergence in monopole differentiates the physical mechanisms behind the two terms.

\section{Iso-qr Integration}\label{sec:qr_integration}
In this appendix, we explain our numerical calculation of integrals of the form
\begin{align}
    \label{eq:integral}
    \int_{r_\mathrm{min}}^{r_\mathrm{max}}
    f(r,q)\,j_\ell(qr)\,,
\end{align}
where $f(r,q)$ is cheap to calculate, but $j_\ell(qr)$ is expensive. This is
the same algorithm, but a more detailed description of what was done in
Sec.~IV of Ref.~\cite{23Gebhard_SFB_eBOSS}. We refer to this algorithm as the
Iso-qr integration. We will first motivate why other approaches are slow or why we
have not considered them, yet, then we derive the requirements for our approach
to work. Finally, we go beyond our initial implementation
\cite{23Gebhard_SFB_eBOSS} and add a restarted-integration that significantly
reduces the number of sampling points needed when the integration interval is
large.

Fundamentally, the approach detailed here is a brute-force trapezoidal
integration. Crucial is our assumption that in \cref{eq:integral} the function
$f(r,q)$ is fast to evaluate, while $j_\ell(qr)$ is expensive.
This allows for $f(r,q)$ to change frequently (hence the need for
efficient evaluation of \cref{eq:integral}), while the spherical Bessel
$j_\ell(qr)$ will be quite static and can be precomputed.
However, the speed of our brute-force integration comes from an informed choice
of the integration nodes, which allows the cache of precomputed $j_\ell(qr)$ to
be very small.

\subsection{Motivation}
\label{sec:qr_despair}
Before we go into detail of our integration method, we review other (often
similar) approaches.

\paragraph{Precompute:}
A simple speedup of the integration can be achieved by precomputing
the spherical Bessels. However, this means creating an array of
$N=N_d\,N_\ell\,N_q\,N_r \simeq 4\cdot200\cdot1000\cdot1000$ floating point
numbers, or about 6~GB of storage, where $N_d$ is the number of derivatives
needed on the spherical Bessel functions, $N_\ell$ is the number of $\ell$, $N_q$ is the
number of sampling points in $q$ that are desired, and $N_r$ is the number of
sampling points in $r$ that are needed for reasonably accurate evaluation of
the integral. For a large redshift range these memory requirements can easily
be a factor of a few larger. It takes both time and memory to precompute all of
this, and leaves little \HL{memory} for other tasks such as the window convolution.

\paragraph{Spline:}
An alternative would be to create splines of the spherical Bessels. The storage
here would be reduced to a 2D array, since the argument can be sampled linearly
from $(qr)_\mathrm{min}=q_\mathrm{min}r_\mathrm{min}$ to
$(qr)_\mathrm{max}=q_\mathrm{max}r_\mathrm{max}$. However, splines generally
require many if-statements with unpredictable branching behavior especially
during the hunting phase on each call. This means spending several tens of CPU
cycles on each grid point.

\paragraph{FFTLog:}
\cref{eq:integral} is almost in a form suitable for the FFTLog
algorithm \citep{00Hamilton_fftlog}. However, FFTLog has two problems: first,
it is not a robust algorithm, and requires apodization and zero-padding when
the integration is over a finite interval. Second, FFTLog requires the function
$f(r,q)$ to be separable so that the $q$-dependence can be taken outside the
integral. Therefore, should $f(r,q)$ be separable at all, it would require some
additional work to expand $f(r,q$) in separable terms. In any case, the first
problem calls for a more robust integration scheme that can serve as a reference
for a possible FFTLog implementation in the future.

\subsection{Iso-\texorpdfstring{$qr$}{qr} lines}
\label{sec:qr_desire}
As mentioned in \cref{sec:qr_despair},
the expensive function $j_\ell(qr)$ only depends on the product $qr$, not on
$q$ and $r$ individually. Therefore, only a 2D array with dimensions in $\ell$
and $x=qr$ need be precomputed on some grid points $(x_m,\ell_k)$.

\paragraph{Neighboring points:}
We can find discretizations of $q$ and $r$ such that all $qr$ fall on
a few iso-$qr$ lines, as follows. Let $q_i$ and $r_j$ be discrete points. Then,
we desire that there are neighboring grid points that fall on the same iso-$qr$
line. For example, if we go up one grid point from $q_i$ to $q_{i+1}$, then we
will need to go down in $r$ from $r_i$ to $r_{i-1}$. For the product to be the
same, we require that
\begin{align}
    \label{eq:qr_jump}
    x_m
    &= q_i\,r_j
    = q_{i+1}\,r_{j-1}\,.
\end{align}
\cref{eq:qr_jump} implies that
\begin{align}
    \label{eq:dlnq_dnlr}
    d\ln q
    = \ln\frac{q_{i+1}}{q_i}
    = \ln\frac{r_j}{r_{j-1}}
    = d\ln r\,.
\end{align}
That is, for points on an iso-$qr$ line, the logarithmic spacings of $q$ and $r$ must have the same step size, or $d\ln q = d\ln r$.

\paragraph{Many points:}
Since we also wish to reuse the same grid points $(q_i,r_j)$ for other iso-$qr$
lines, it means that the spacing for both $q$ and $r$ needs to be logarithmic
throughout the entire integration plane. The mapping from $(i,j)$ to the index
$m$ in \cref{eq:qr_jump} is simply $m=i+j-1$.

For the integration in log-space, we choose a simple mid-point trapezoidal
rule, which will be simplest when the number of subdivisions is an integer. If
$N_r$ is the number of desired subdivisions in $r$, then
\begin{align}
    R = d \ln r = \frac{\ln r_\mathrm{max} - \ln r_\mathrm{min}}{N_r}\,,
\end{align}
and the corresponding grid points are
\begin{align}
    r_j = r_\mathrm{min} \, e^{(j-\frac12) R}\,,
\end{align}
where $j=1,2,\ldots,N_r$. The factor $e^{\frac12R}$ ensures that the
grid points fall in the middle of each interval.

\subsection{Striding}
\label{sec:qr_hope}
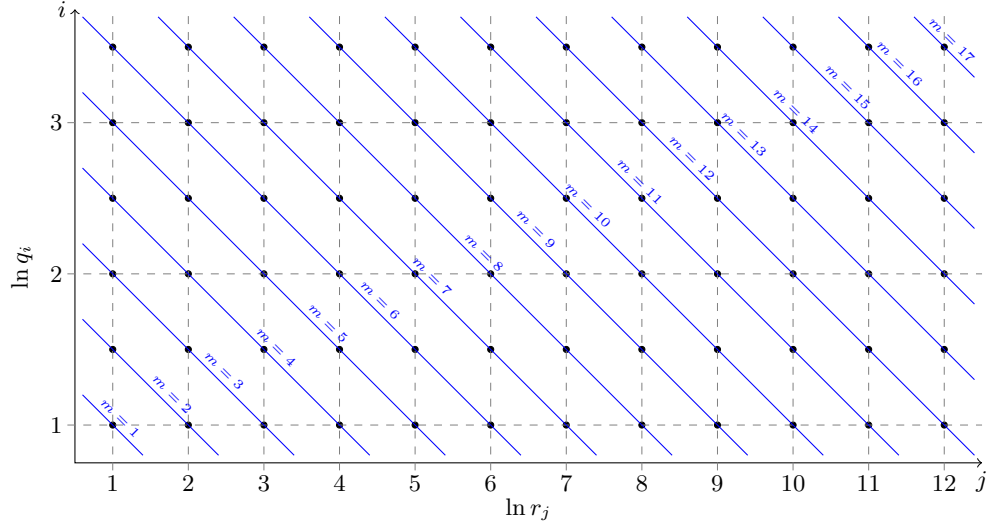
\begin{figure*}
    \centering
    \centerline{\begin{tikzpicture}[scale=1]
        \tikzmath {
            \qstride = 2;
            \qmax = 3;
            \imax = \qstride * \qmax;
            \rstride = 1;
            \rmax = 12;
            \jmax = \rstride * \rmax;
            \mmax = \imax + \jmax - 1;
        }
        \draw[<->] (0,\imax) node[left]{$i$} -- node[left,xshift=-20,rotate=90]{$\ln q_i$} (0,0)
            -- node[below,yshift=-10]{$\ln r_j$} (\jmax,0) node[below]{$j$};
        \foreach \r in {1,2,...,\jmax} {
            \foreach \q in {1,2,...,\imax} {
                \draw[fill] (\r-0.5,\q-0.5) circle[radius=0.04];
            }
        }
        \foreach \i in {1,2,...,\qmax} {
            \draw (-0.05,\qstride*\i-\qstride+0.5) node[left]{\small $\i$};
        }
        \foreach \i in {1,2,...,\rmax} {
            \draw (\rstride*\i-\rstride+0.5,-0.05) node[below]{\small $\i$};
        }
        \foreach \i in {1,2,...,\qmax} {
            \draw[dashed,help lines] (-0.1, \qstride*\i-\qstride+0.5) -- (\jmax, \qstride*\i-\qstride+0.5);
        }
        \foreach \i in {1,2,...,\rmax} {
            \draw[dashed,help lines] (\rstride*\i-\rstride+0.5,-0.1) -- (\rstride*\i-\rstride+0.5,\imax);
        }
        \foreach \m in {1,2,...,\mmax} {
            \tikzmath {
                \Ax = max(0, \m - \imax) + 0.1;
                \Ay = min(\m, \imax) - 0.1;
                \Bx = min(\m, \jmax) - 0.1;
                \By = max(0, \m - \jmax) + 0.1;
            };
            \draw[blue] (\Ax,\Ay) -- (\Bx,\By);
        }
        \foreach \m in {1,2,...,\mmax} {
            \tikzmath {
                \x = (\jmax - 1) * (\m - 1) / (\mmax - 1) + 0.5 + 0.1;
                \y = (\imax - 1) * (\m - 1) / (\mmax - 1) + 0.5 + 0.1;
            }
            \draw (\x,\y) node[rotate=-45]{\color{blue}{\tiny $m=\m$}};
        }
    \end{tikzpicture}}
    \caption{
        The $qr$ grid in log-space. Solid lines show the $qr=\mathrm{const}$
        lines, dashed lines show possible integration paths, here with skipping
        some of the $q$ points. The axis labels refer to $i$ and $j$,
        respectively.
    }
    \label{fig:qr_grid_log}
\end{figure*}

\paragraph{Stride:}
It is our desire to use as few grid points as possible. Indeed, we can skip
some of the $q_i$. If we request $\tilde N_q$ grid points in $q$, then our
requirement of using the same logarithmic spacing as in $r$ means that we need
to skip $\tilde s_q-1$ grid points, we call $\tilde s_q$ the \emph{stride}. Its
value is given by
\begin{align}
    \tilde s_q &= \frac{\ln q_\mathrm{max} - \ln q_\mathrm{min}}{\tilde N_q\,d\ln r}\,.
\end{align}
To gain some intuition for this equation, imagine $\tilde N_q=1$. Then, we
should skip all but one $q$ grid point, so $\tilde s_q$ is very large. If, on
the other hand, $\tilde N_q$ happens to be the number of logarithmic
subdivisions of size $d\ln r$, then we shouldn't skip any, and our stride is
$\tilde s_q=1$.

\paragraph{Integer stride:}
Of course, the actual stride $s_q$ must be an integer, and we
round $\tilde s_q$ down to obtain it. That also means that the actual number
$N_q$ of grid points cannot be equal to the desired number $\tilde N_q$ of grid
points, or $N_q\neq\tilde N_q$ in general. Similarly, only one of $q_{\rm min}$
and $q_{\rm max}$ can be exactly included as a grid point. In case the exact
interval is needed, one (expensive) extra step will need to be included in the
integration.

\cref{fig:qr_grid_log} illustrates the grid with $s_q=2$. A stride $s_r$ in $r$
can also differ from unity, and we allow for strides in both axes below.
Important for the implementation is how to map $(i,j)$ to $m$. We use
\begin{align}
    \label{eq:qr_m}
    m &= 1 + (i - 1) \, s_q + (j - 1) \, s_r \,,
\end{align}
for $i=1,\ldots,N_q$, $j=1,\ldots,N_r$, and $m=1,\ldots,N_{j_l}$, where
$N_{j_l}$ is the number of spherical Bessels that need to be stored.

\subsection{Sampling needs}
\label{sec:qr_vici}
To get a reasonably accurate estimation of \cref{eq:integral}, the numerically
calculated integrand needs to sufficiently sample the oscillations of the
spherical Bessel functions. We only use the spherical Bessel in this step,
because we assume that the function $f(r,q)$ is smooth and varies on the same
scales or slower.

\paragraph{Sampling spherical Bessels:}
Since the spherical Bessels asymptote to $j_\ell(x)\to\sin(x+\phi_\ell)/x$ as
$x\to\infty$, and the frequency of oscillations is only lower towards $x\to0$,
we desire a maximum $\Delta x$ of
\begin{align}
    \Delta x \leq \frac{2\pi}{n_\mathrm{samp}}\,,
\end{align}
where $n_\mathrm{samp}$ is the number of sampling points in one oscillation.
Due to the logarithmic spacing, the largest $\Delta x_m$ will be given at
$x_\mathrm{max}=q_\mathrm{max}r_\mathrm{max}$, and this informs the total
number of sampling points $N_r$ and $N_q$.

\subsection{Interval Integration}
\label{sec:interval_integration}
The logarithmic sampling required by our $qr$-integration method can lead to a
very fine sampling near the lower integration bound, e.g., if $r_{\min}\sim0$,
as occurs for the integrated GR terms. In this case, we can divide the
integration range into smaller intervals, and apply the method on each interval
individually. The subdividing can be done recursively until the lower bound is
smaller than the largest step. In practice, this buys about a factor \HL{of} two in
speedup.

Crucial is that the $q$ array remains the same across each interval. That is, we
first choose a $q$ array, and then we will choose the individual $r$ arrays in
each interval.\footnote{As a side effect this will also allow a simple
implementation for cross-correlations of tracers with different $r_{\min}$ and
$r_{\max}$.}
If the logarithmic spacing in $q$ is $Q=\dd\ln q$, and the  spacing in $r$ is
$R_n=\dd\ln r_n$ at the $n^{\rm th}$ interval, our integration technique requires that
\begin{align}
    \frac{R_n}{s_r^n} &= \frac{Q}{s^n_q}\,,
    \label{eq:qr_stridesizes}
\end{align}
where $s^n_r$ and $s^n_q$ are positive integers. $s^n_q$ is the stride in $q$
relative to a fictitious $\dd\ln r=Q/s^n_q$, and $s^n_r$ is the stride in $r$
relative to that same fictitious $\dd\ln r$.

If $N_r^n$ is the number of integration nodes in interval $n$, then the width
of the interval in log space is $N^n_r \, R_n = N_r^n \, s_n \, Q$, and
we defined $s_n\equiv s^n_r/s^n_q$. All intervals together must add to approximately
the full interval,
\begin{align}
    \label{eq:qr_lnrmax_lnrmin}
    \ln r_{\max} - \ln r_{\min}
    \geq
    Q\sum_n N^n_r \, s_n\,,
\end{align}
where the inequality will need to be overcome by an extra integration step
discussed below.
The maximum $\ln r$ in each interval is
\begin{align}
    \label{eq:qr_an_bn}
    \ln b_n &= \ln r_{\max} - Q\sum_{i=n+1}^{n_{\max}} N_r^i\,s_i \,,
\end{align}
so that the $n^\mathrm{th}$ interval ranges from $b_{n-1}$ to $b_n$, and
intervals are labeled by counting from the bottom, i.e., the interval $n=1$ has lower
bound $\simeq r_{\min}$, and the interval $n_{\max}$ as upper bound $r_{\max}$.

The precision of the integration is \HL{controlled} by requiring that the maximum
linear integration step is below some $\Delta r_{\max}$ for every interval,
\begin{align}
    \label{eq:qr_drmax}
    \Delta r_{\max}
    \geq
    b_n R_n
    =
    \frac{r_{\max}\,s_n\,Q}{\exp(Q\sum_{n+1}^{n_{\max}} N^n_r\,s_n)}\,.
\end{align}
Solving for $s_n$ we obtain,
\begin{align}
    s_n
    &\leq
    \frac{\Delta r_{\max}}{Q\,r_{\max}}
    \,\exp(Q\sum_{n+1}^{n_{\max}} N^n_r\,s_n)\,,
\end{align}
which is a series of equations for each $n$. When $n=n_{\max}$ this gives a
maximum value for $s_{n_{\max}}\leq\frac{\Delta r_{\max}}{Q\,r_{\max}}$, and
using this value will maximize the step size in the upper-most interval.


$s_{n_{\max}}$ in turn can be used to calculate the minimum $N_r^{n_{\max}}$,
provided that $s_{n_{\max}-1}$ is given, or, more generally,
\begin{align}
    N^n_r
    &\geq
    \frac{1}{s_n\,Q}\ln\frac{s_{n-1}}{s_n}\,,
\end{align}
which requires all the $s_n$ to be given. Using \cref{eq:qr_lnrmax_lnrmin} we
get the constraint
\begin{align}
    \ln\frac{r_{\max}}{r_{\min}}
    &\geq
    \sum_n \ln\frac{s_{n-1}}{s_n}\,,
\end{align}
which will need to be checked so as not to overshoot $r_{\min}.$

The last step from slightly above $r_{\min}$ to $r_{\min}$ can be done with a
single step of arbitrary width such that the full integration interval from
$r_{\min}$ to $r_{\max}$ is included.

The goal, now is to find integer strides $s_q^n\geq1$ and $s_r^n\geq1$ that satisfy these
constraints with $s_n=s_r^n/s_q^n$, and that minimize the size of the $qr$-qr
and $j_l$-cache.
For a given $\ell$, the number of grid points is given by
\begin{align}
    N_{qr}^\mathrm{total}
    &=
    N_q \sum_n N^n_r
    \,,
\end{align}
and the number of spherical Bessel evaluations is [cf.~\cref{eq:qr_m}]
\begin{align}
    N_{j_l}^\mathrm{total}
    &=
    N_d \sum_n \left[ 1 + (N_q-1) s^n_q + (N_r^n-1) s^n_r \right],
\end{align}
where $N_d$ is the number of derivatives of the spherical Bessels that are
needed. We then minimize some combination $F(s^n_q, s^n_r) =
N_{qr}^\mathrm{total} + \beta N_{j_l}^\mathrm{total}$ for some $\beta$.
The choice $\beta=1$ is natural as it will minimize the number of memory
locations that need to be accessed. 
We have now generalized the Iso-qr integration to multiple subintervals of $r$, while ensuring the $q$ array to remain the same across the entire interval.

\section{Useful Identities}

\noindent\textbf{Orthogonality of spherical harmonics:}
\be
\label{eq:YY_ortho}
    \int d^2 \hat{\bk}~ Y_{\ell m} (\hat{\bk}) Y^*_{\ell' m'}(\hat{\bk}) = \delta_{\ell \ell'} \delta_{m m'}\,.
\ee
\noindent\textbf{Rayleigh expansion of a plane wave:}
\begin{equation}
    e^{i \bk \cdot \bx} = (4\pi)\sum_{\ell m} i^\ell j_\ell(kx) Y_{\ell m} (\hat{\bx}) Y_{\ell m}^* (\hat{\bk})\,.
\label{eq:planewave-Ylm}
\end{equation}

\noindent\textbf{Orthogonality of spherical Bessel functions}:
\begin{equation}
\label{eq:O_jl}
\int_{r}r^2j_{\ell}(kr)j_{\ell}(k'r)=\frac{\pi}{2kk'}\delta^{{\rm D}}(k-k').
\end{equation}

\bibliography{refs}

\end{document}